\documentclass[useAMS,usenatbib]{mnras}

\usepackage{newtxtext,newtxmath}
\usepackage[T1]{fontenc}
\usepackage{ae,aecompl}

\usepackage{verbatim}
\usepackage{natbib}
\usepackage{amsfonts}
\usepackage[usenames,dvipsnames]{color}
\usepackage[normalem]{ulem}

\usepackage{graphicx}	% Including figure files
\usepackage{amsmath}	% Advanced maths commands
\usepackage{amssymb}	% Extra maths symbols

\newcommand{\be}{\begin{equation}}
\newcommand{\ee}{\end{equation}}
\newcommand{\bes}{\begin{equation*}}
\newcommand{\ees}{\end{equation*}}
\newcommand{\bea}{\begin{eqnarray}}
\newcommand{\eea}{\end{eqnarray}}
\newcommand{\beas}{\begin{eqnarray*}}
\newcommand{\eeas}{\end{eqnarray*}}

\newcommand{\msun}{M_{\odot}}

\newcommand{\mpch}{\;{\rm Mpc}/h}
\newcommand{\de}{{\rm d}}

\newcommand{\dstheta}{\Delta\Sigma^{4\theta}}
\newcommand{\dsconst}{\Delta\Sigma^{\rm const}}
\newcommand{\dsthetax}{\Delta\Sigma^{4\theta}_\times}
\newcommand{\dsconstx}{\Delta\Sigma^{\rm const}_\times}

\newcommand{\pmem}{p_{\rm mem}}

\newcommand{\nsat}{N_{\rm sat}}
\newcommand{\redmap}{redMaPPer\,}
\newcommand{\redmagic}{redMaGiC\,}

\title[Cluster Ellipticity]{The ellipticity of galaxy cluster halos from satellite galaxies and weak lensing}

\author[Shin et al.]
{\parbox{\textwidth}{
  Tae-hyeon~Shin$^{1}$\thanks{E-Mail: taeshin@sas.upenn.edu},
  Joseph~Clampitt$^{1}$,
  Bhuvnesh~Jain$^{1}$,\\
  Gary Bernstein$^{1}$,
  Andrew~Neil$^{1,2}$,
  Eduardo~Rozo$^{3}$,
  Eli~Rykoff$^{\,4}$} \\
  \vspace{0.2cm}\\
  $^1$Department of Physics and Astronomy, Center for Particle Cosmology, \\
  \;University of Pennsylvania,
  209 S. 33rd St., Philadelphia, PA 19104, USA \\
  $^2$Department of Astronomy and Astrophysics, University of Chicago, Chicago, IL 60637 \\
  $^3$Department of Physics, University of Arizona, 1118 E 4th St, Tucson, AZ 85721 USA \\
  $^4$SLAC National Accelerator Laboratory, Menlo Park, CA 94025 USA}

\date{Accepted XXX. Received YYY; in original form ZZZ}

\pubyear{2017}

\begin{document}
\label{firstpage}
\pagerange{\pageref{firstpage}--\pageref{lastpage}}
\maketitle

\begin{abstract}
We study the ellipticity of galaxy cluster halos as characterized by the distribution of cluster galaxies and as measured with weak lensing. We use monte-carlo simulations of elliptical cluster density profiles to estimate and correct for Poisson noise bias, edge bias and projection effects.
We apply our methodology to 10,428 SDSS  clusters identified by the \redmap algorithm with richness above 20.
We find a mean ellipticity $= 0.271 \pm 0.002$ (stat) $\pm 0.031$ (sys) corresponding to an axis ratio $= 0.573 \pm 0.002$ (stat) $\pm 0.039$ (sys).
We compare this ellipticity of the satellites to the halo shape, through a stacked lensing measurement using optimal estimators of the lensing quadrupole based on Clampitt and Jain (2016).
We find a best-fit axis ratio of $0.56 \pm 0.09$ (stat) $\pm 0.03$ (sys),
consistent with the ellipticity of the satellite distribution. Thus cluster galaxies trace the shape of the dark matter halo to within our estimated uncertainties.
Finally, we restack the satellite and lensing ellipticity measurements along the major axis of the cluster central galaxy's light distribution.
From the lensing measurements we infer a misalignment angle with an RMS of  ${30^\circ \pm 10}^\circ$ when stacking on the central galaxy. We discuss applications of halo shape measurements to test the effects of the baryonic gas and AGN feedback, as well as dark matter and gravity. The major improvements in signal-to-noise expected with the ongoing Dark Energy Survey and future surveys from LSST, Euclid and WFIRST will make halo shapes a useful probe of these effects.
\end{abstract}


%%%%%%%%%%%%%%%%%%%%%%%%%%%%%%%%%%%%%%%%%%%%%%%%%%

%%%%%%%%%%%%%%%%% BODY OF PAPER %%%%%%%%%%%%%%%%%%

%%%%%%%%%%%%%%%
\section{Introduction}
%%%%%%%%%%%%%%%

Since dark matter contributes more mass than gas or stars to the halos of galaxies and galaxy clusters, it has a strong influence on the formation of visible galaxies.
This means the correlation between dark matter and visible galaxy properties -- such as mass, shape, and alignment -- can provide constraints on theories of galaxy formation.
In practice, though, such correlations are challenging. 

However, galaxy clusters, composed of many satellite galaxies orbiting within a common dark matter halo, provide a unique solution.
While this cluster halo is again not directly visible, cluster satellites are expected to trace the halo's shape.
Thus, correlations between the brightest cluster galaxy (BCG), or central galaxy (CG), its major axis and the locations of cluster satellites provide valuable information.
Early attempts to detect this correlation were made by \citet{kas02} with a small cluster sample.
The first detection of alignment between the BCG major axis and cluster satellites was made by \citet{brainerd05}.
Previous results were noisy but showed hints of anti-alignment.
Later work by \citet{yang06} confirmed the major axis alignment of \citet{brainerd05}.
\citet{falt07}, \citet{azzaro07}, and \citet{wang08} found further confirmation using the larger SDSS DR4 data set.
These studies were based on rather small groups with typically a few members per group.
This was followed by \citet{falt08} which used N-body simulations to explain the same measurements.

\citet{agustsson06} and \citet{afp06} also used N-body simulations to explain the alignments found in data.
\citet{kang07} improved on this work with more detailed simulations and mock galaxy catalogs, including using a semi-analytic model of galaxy formation to place galaxies in N-body halos.
\citet{kang07} found that aligning galaxies perfectly with dark matter halos over-predicted the observed alignment signal.
Thus some misalignment would be necessary to explain the data.

\citet{wang08} compared measured alignments to Monte Carlo simulations of both triaxial (using the \citealt{jing02} model) and isotropic halos.
These simulations allowed \citet{wang08} to take into account discrete sampling issues due to the small number of satellite galaxies.
The data strongly preferred triaxial models which are projected to 2D elliptical models and furthermore allowed \citet{wang08} to fit for the halo axis parameters.
\citet{nsd10} confirmed the significant correlation between BCG and satellite alignment using SDSS clusters with more than 20 members.
They also discerned an $\sim 4 \sigma$ trend with BCG dominance: clusters with larger differences in brightness between the brightest and 2nd and 3rd brightest members had a stronger BCG-satellite alignment.

Recently, \citet{huang16} carried out a methodical study of SDSS \redmap clusters. They measured a 35$^\circ$ average offset between central galaxy (CG) major axis and major axis of satellite distribution.
In addition, they found additional correlations between several cluster properties and alignment.
Although these correlations are weak -- roughly $0.1$ for central dominance, absolute magnitude, central size, and centering probability -- they were detected at $10 \sigma$ significance.
This is made possible by the large size of the SDSS DR8 data.

The \citet{huang16} measurements are state-of-the-art for cluster alignment studies.
However, the physical interpretations presented in \citet{huang16} still need to assume that cluster satellite galaxies are an unbiased probe of the underlying dark matter.
While this may be true qualitatively, effects such as Poisson sampling bias (which is explained in this paper later on) can cause an offset between the dark matter major axis and the major axis of the satellite distribution.
This is especially limiting for lower richness clusters since smaller number of satellite galaxies causes the larger offset.
Even for richer clusters, where Poisson noise is less problematic, a direct probe of the dark matter would complement the information from visible satellites.

Gravitational lensing provides such a direct probe of the dark matter halos.
Early attempts by \citet{hoekstra04, parker07} using the first quadrupole estimators were not able to detect the effect.
\citet{mhb06}, with the quadrupole estimators of \citet{nr2000}, were somewhat more successful.
\citet{mhb06} also used the larger SDSS data set, probing alignments of galaxies and dark matter with several colour and luminosity bins.
They found the strongest alignment for the brightest, red lenses, with a significance $\sim 2-3\sigma$.
Other work by \citet{van12} and \citet{schrabback15} with the same estimator and other surveys found little alignment.

Lensing ellipticity measurements on cluster and group scales have been more successful.
\citet{oguri10} studied the ellipticity of $\sim 20$ massive clusters using an elliptical NFW fits to individual cluster shear profiles.
They reported a $\sim 7\sigma$ detection of cluster ellipticity for a subset of the clusters that were well fit by the elliptical NFW profile.
With a larger sample of 4300 clusters, \citet{evans09} obtained a $\sim 3.5 \sigma$ detection of stacked cluster ellipticity.
Note that \citet{oguri10} has a much  higher number density for the source galaxies than \citet{evans09} does.
More recently \citet{van16} reported a $3-4 \sigma$ detections of halo ellipticity of GAMA groups with KiDS shear catalogs.
\citet{van16} used several methods for stacking the groups, including aligning with the cluster BCG and the major axis of group members.

Previously, in \citet{clampitt16} we measured the halo ellipticity of Luminous Red Galaxies (LRGs, $4\sigma$ significance) and \redmap clusters ($3\sigma$) using new estimators.
These estimators involved shears, $\gamma_1$ and $\gamma_2$, in a Cartesian coordinate system defined by aligning the major axes of all lenses (LRGs) in the stacked measurement.
We take the estimator of \citet{clampitt16} one step further and derive the optimal halo ellipticity estimators.
The quadrupole naturally has two optimal estimators: one measures a $4\theta$ angular variation in background shears, and the other measures a constant-angle signal.
These estimators are subject to different systematics, providing natural checks of our measurement.

In addition, we stack the lensing measurement using the major axis of the CG light profile, as well as the major axis of the satellite distribution.
These two ways of stacking are also subject to different systematics, providing further cross-checks.
Our results are robust: all four methods (2 estimators and 2 ways of aligning lenses) of measuring halo ellipticity for these \redmap clusters are in agreement.

In section 2 we describe the clusters and lensing shear data used in this work.
Section 3 presents our methodology and results for SDSS cluster ellipticity of satellite distribution.
Section 4 defines the optimal quadrupole lensing estimators and presents the results of the measurement on SDSS clusters.
Section 5 compares our results to other work.
In section 6 we discuss the results and their implications.

%%%%%%%%%%%%%%%
\begin{figure*}
\centering
\resizebox{58mm}{!}{\includegraphics{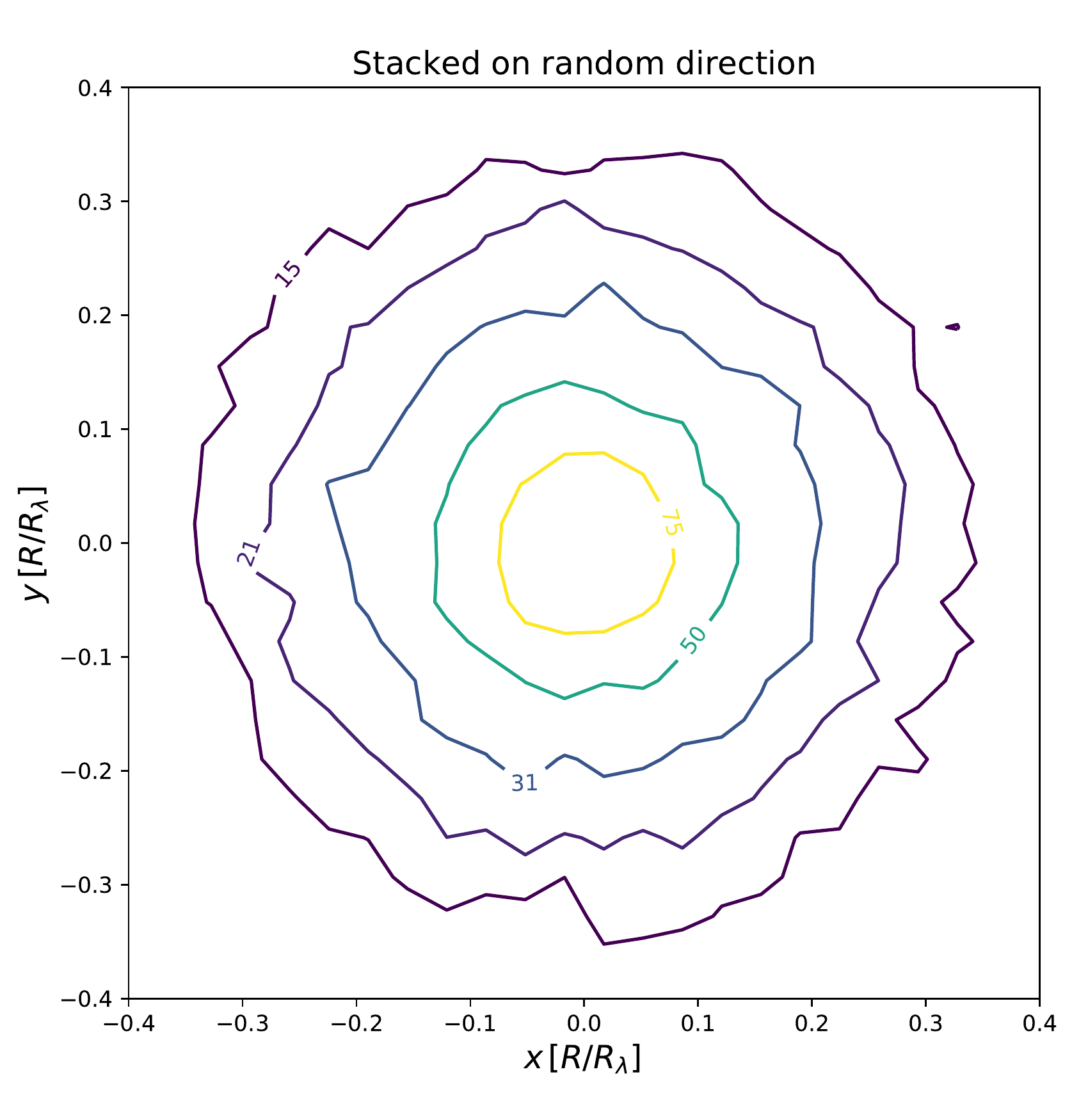}}
\resizebox{58mm}{!}{\includegraphics{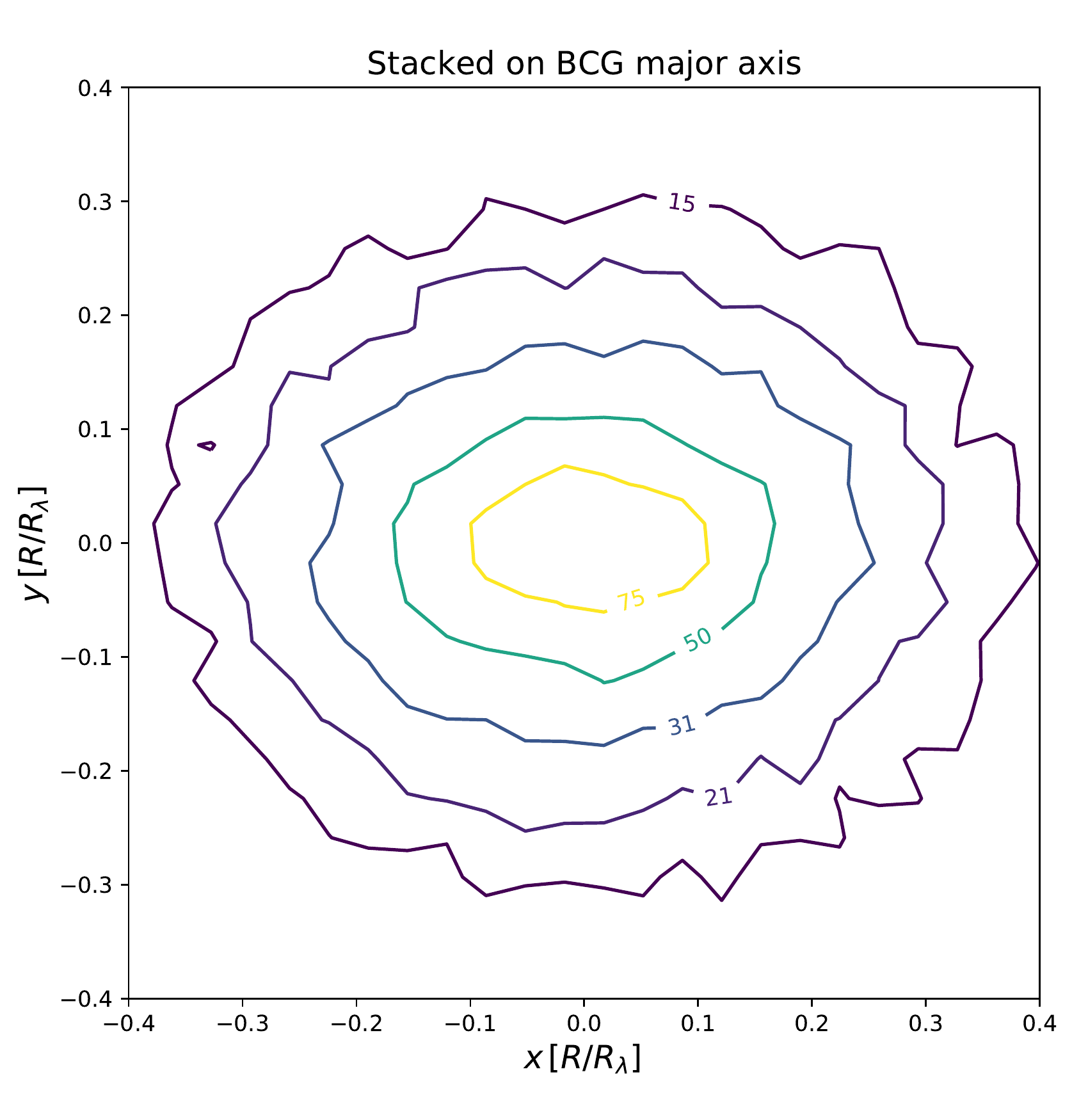}}
\resizebox{58mm}{!}{\includegraphics{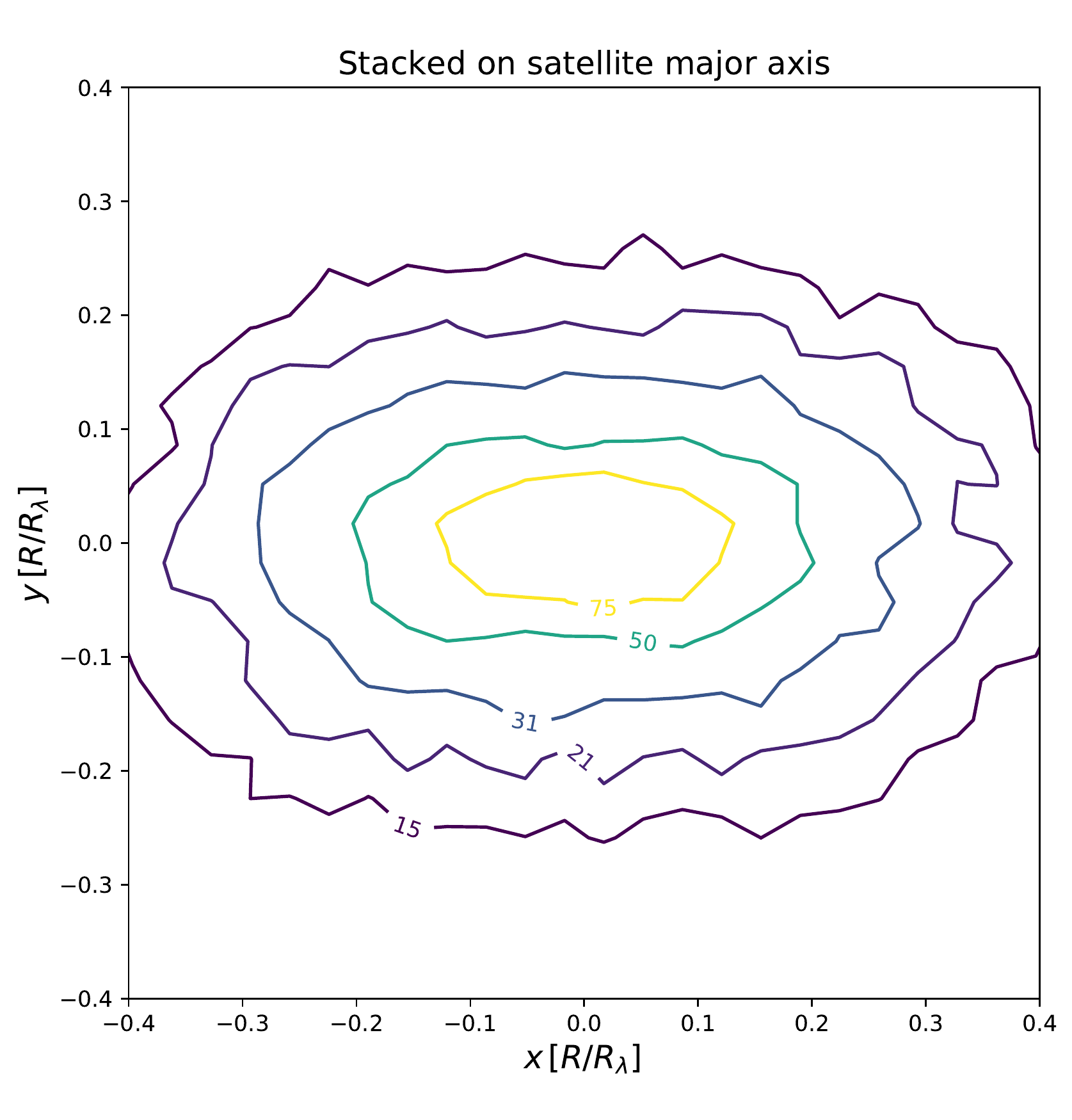}}
\caption{Satellite density contours of all clusters, with random orientations (left panel), stacked to align the major axis of the central galaxy (middle panel), and stacked along the major axis of the satellite distribution itself (right panel), in the unit of ``galaxy number density per ${\rm Mpc^{2} h^{-2}}$ per cluster''. Comparing the middle and right panels shows how the satellite distribution in the outer parts is misaligned with respect to the central galaxy. The sample is restricted to the $\sim$ 6,700 subset with a measured major axis for the central galaxy.
}
\label{fig:sat-contour}
\end{figure*}
%%%%%%%%%%%%%%%%%%%%%%%

%%%%%%%%%%%%%%%
\section{Data}
%%%%%%%%%%%%%%%

%%%%%%%%%%%%%%%
\subsection{redMaPPer Clusters in SDSS}
\label{sec:rm-data}
%%%%%%%%%%%%%%%

We use the public SDSS DR8 \redmap cluster catalog presented in \citet{rozo15a} (See also \citealt{rykoff14} for more details of the algorithm).

Using red galaxies having luminosity $L(z) \geq L_*(z)$, where $L_*$ is the characteristic luminosity defined in \citet{rykoff14} (see their Sec. 4), \redmap assigns each cluster a richness, $\lambda$, which quantifies the probable number of cluster satellites. 
$\lambda$ is given as
\be
\lambda = \sum_i p_{{\rm mem},i} = \sum_i \frac{\lambda u(\textbf{x}_{i}|\lambda)}{\lambda u(\textbf{x}_{i}|\lambda) + b(\textbf{x}_{i})} \, .
\ee
Note that $\lambda$ is the only unknown in the equation, which can be numerically solved.
Here, $u(\textbf{x}_{i}|\lambda)$ is the normalized density profile of the cluster and $b(\textbf{x}_{i})$ is the density of the uniform background, while $\textbf{x}_{i}$ is a vector that represents the projected radius, the luminosity and the colour of the $i$-th candidate member galaxy (See \citet{rykoff14} for details on how they model $u(\textbf{x}|\lambda)$ \& $b(\textbf{x})$ and determine $\lambda$).
Note that we round the richness value to the closest integer when performing MC simulations (Sec.~\ref{sec:satellite}).

We use all clusters in the range  $20 < \lambda < 200$ and with redshifts between $0.1 < z < 0.41$.
The upper redshift limit is necessary to ensure sufficiently many background sources for lensing measurements.
For the $i$-th potential cluster central galaxy (CG), \redmap calculates a centering probability, $P_{{\rm cen},i}$ as follows: 
\be
p_{\rm cen}(\textbf{y}_{i}) = p_{\rm free}(\textbf{y}_{i}) \frac{u_{\rm cen}(\textbf{y}_{i})}{u_{\rm cen}(\textbf{y}_{i}) + (\lambda-1) u_{\rm sat}(\textbf{y}_{i}) + u_{\rm bg}(\textbf{y}_{i})} \, \\
\ee
\be
P_{{\rm cen},i} \propto p_{\rm cen}(\textbf{y}_{i}) \prod_{j \neq i} (1-p_{\rm cen}(\textbf{y}_{i}))\, ,
\ee
with a normalization condition
\be
1 = \sum_i P_{{\rm cen},i} \, .
\ee
Note that $P_{{\rm cen}}$ and $p_{{\rm cen}}$ are different from each other. 
Here, $p_{\rm free}$ is the probability that a galaxy is partially masked by another cluster.
$u_{\rm cen}$, $u_{\rm sat}$ and $u_{\rm bg}$ are the normalized distributions of central, satellite and background galaxies, respectively, where $\textbf{y}_{i}$ is a vector containing magnitude, photo-z and local galaxy density of the $i$-th galaxy (see \citet{rykoff14} for details on the filters they apply to model $u_{\rm cen}$, $u_{\rm sat}$ and $u_{\rm bg}$).

We only use \redmap clusters for which the CG is unambiguous, i.e., clusters with $P_{\rm cen} > 0.9$.
We have tested and confirmed that the cut on $P_{\rm cen}$ does not alter our fiducial results significantly.

%%%%%%%%%%%%%%%
\subsection{Galaxy shapes}
\label{sec:data-shape}
%%%%%%%%%%%%%%%

For lensing measurements we use an SDSS shear catalog with 34.5 million sources \citet{sjs2009}.
For each source we also have a photometric redshift probability distribution, described in \citet{scm12}, which is necessary to apply the optimal lensing weight to each lens-source pair.
The total redshift distribution peaks at $z \sim 0.35$ with a tail towards higher redshifts.

We also use the shear catalog to find the major axes of the CG light distributions.
This will be necessary to align our Cartesian coordinate systems for each lens prior to performing a stacked lensing measurement.
The shear is estimated with the second moment method with an elliptical gaussian weight that is iteratively readjusted to the size and shape of the object, with Petrosian radius and photometric centroid as the initial guesses for the size and position of the source (see \citet{sjs2009} for details).
We match the shear catalog of \citet{sjs2009} to all CG in the \redmap catalog.
This results in 6,681 matching CGs (64\% of CGs have a successful match) in our lens sample. 
We have confirmed that these 6,681 clusters are an unbiased subsample of the full sample
(see Sec.~\ref{sec:cg-stack} for the details of this test).
Essentially all of these matched CGs have quality shear measurements, based on the small shape measurement error of each galaxy.
We calculate the orientation of CGs and its uncertainty using: $\tan{2\phi} = e_2/e_1$, where $\phi$ and $e_i$ represent the orientation angle and the shear components of the object. The distribution of uncertainty of the orientation angles peaks at zero and 68\% (95\%) of them are smaller than $1.65^\circ$ ($5.75^\circ$).
See \citet{sjs2009} for more details on the shear catalog, and \citet{sjf2004} for more detailed descriptions of the shear measurement method.

Throughout the paper we use a spatially flat $\Lambda$CDM cosmology with $\Omega_m = 0.3$ and show distances in {\it physical} (not comoving) Mpc/$h$.

%%%%%%%%%%%%%%%
\section{Measurement of Ellipticity of Satellite Distribution}
\label{sec:satellite}
%%%%%%%%%%%%%%%

%%%%%%%%%%%%%%%
\subsection{Raw ellipticity measurement}
\label{sec:method}
%%%%%%%%%%%%%%%

The contours of \redmap satellite density are shown in Fig.~\ref{fig:sat-contour}, with the position re-scaled with the cluster size, $R_{\lambda}$ (see Sec.~\ref{sec:edge-bias}).
The left panel shows the stacked contours with random orientations.
The middle panel shows the result when aligning the stack along CG major axes (estimated as in Sec.~\ref{sec:data-shape}).
The right panel shows the result of aligning along the satellite major axes  (estimated as in Sec.~\ref{sec:method}).
The randomly oriented stack has circular contours.
Even by eye the nonzero ellipticity of the other two panels is clear.
Furthermore, the stack along satellite major axes is visibly more elongated than the stack along CG major axes.
Later we will show this is a systematic elongation due to Poisson noise in the satellite distribution (see Sec.~\ref{sec:bias}).
Such systematics make interpretation of the contour plots difficult.

Given a set of $\lambda$ Cartesian satellite coordinates, e.g., $(x_i, y_i)$ for satellite $i$, we can estimate the ellipticity of the cluster's surface density.
We define the two components of ellipticity as
\be \label{eq:e1}
e_1 = \frac{I_{xx} - I_{yy}}{I_{xx} + I_{yy}}
\ee
\be \label{eq:e2}
e_2 = \frac{2 I_{xy}}{I_{xx} + I_{yy}} \, ,
\ee
where the second moments are given by
\be \label{eq:moments}
I_{xx} = \frac{\sum_i x_i^2 w_i}{\sum_i w_i} \, , \,\,\,\,\,
I_{yy} = \frac{\sum_i y_i^2 w_i}{\sum_i w_i} \, , \,\,\,\,\,
I_{xy} = \frac{\sum_i x_i y_i w_i}{\sum_i w_i} \, ,
\ee
with weights $w_i = 1/(x_i^2 + y_i^2)$.
We define the centre of the coordinates as the centre of each \redmap cluster in the catalog.
Besides, values of the second moments above are weight-dependent. Therefore, the weight can bias the measured ellipticity from the true ellipticity\footnote{The true ellipticity refers to the average ellipticity of the underlying distribution according to which satellites are generated inside  $R_{\lambda}$. Note that we assume that the ellipticity is constant with radius  inside a cluster in our MC simulation. To confirm that our assumption is accurate enough for the purpose of measuring the average ellipticity, we verified in the appendix~\ref{sec:ellip-dist} that our MC simulation retrieves the distribution of the measured ellipticity very well.}. However, we apply the same weight in our Monte Carlo (MC) simulation (see Sec.~\ref{sec:bias}) to properly account for the effect. Therefore, the true retrieved ellipticity would not be biased.
The two components of ellipticity can also be written in terms of the total magnitude
\be \label{eq:emag}
e = \sqrt{e_1^2 + e_2^2} \, ,
\ee
and position angle of the major axis relative to the x-axis
\be
\tan{2\phi} = \frac{2 I_{xy}}{I_{xx} - I_{yy}} \, .
\ee
It will sometimes be useful to convert the magnitude of the ellipticity to an axis ratio $q$, using
\be \label{eq:axis}
e = \frac{1 - q}{1 + q} \, .
\ee

Next we divide the \redmap clusters into richness bins and measure the average ellipticity according to Eqs.~(\ref{eq:e1}) -- (\ref{eq:emag}).
In order to apply these equations, we must choose which potential cluster members to include in the measurement.
For each cluster and each red galaxy, \redmap estimates $\pmem$, the probability that the red galaxy is a member of that cluster.
This membership probability condenses galaxy colour, luminosity, and projected distance from the cluster centre into a number between 0 and 1 for each red galaxy.
In order to minimize the number of possible interlopers (see Sec.~\ref{sec:fake-member}), we need to remove galaxies with small $\pmem$ values.
However, $\pmem$ tends to decrease quickly with increasing projected distance from the cluster centre.
Therefore, if we apply a $\pmem$ cut too aggressively, most of the member galaxies on the outskirts would be filtered out and our ellipticity estimates would only apply to the cluster centre.
To balance these competing effects, we choose an intermediate cut of $\pmem > 0.5$ for our fiducial results.
We have tested other cuts and find that the average ellipticity across all $\lambda$ bins is not sensitive to the precise cut.
See Appendix~\ref{sec:app-pmem} and Sec.~\ref{sec:edge-bias} for a fuller discussion of the effects of $\pmem$ cuts.

On the other hand, in \redmap algorithm, the projection effect due to nearby halos is likely to be aspherical.
For instance, if a projected halo is located to the right of the parent halo, we get an ellipticity enhancement in that direction.
However, this ellipticity enhancement is randomly oriented, so when stacking a large number of clusters, we obtain an effectively circularized distribution of the ``projected'' halos.
It does not change the calculation we make in this paper.

\begin{figure}
\centering
\resizebox{85mm}{!}{\includegraphics{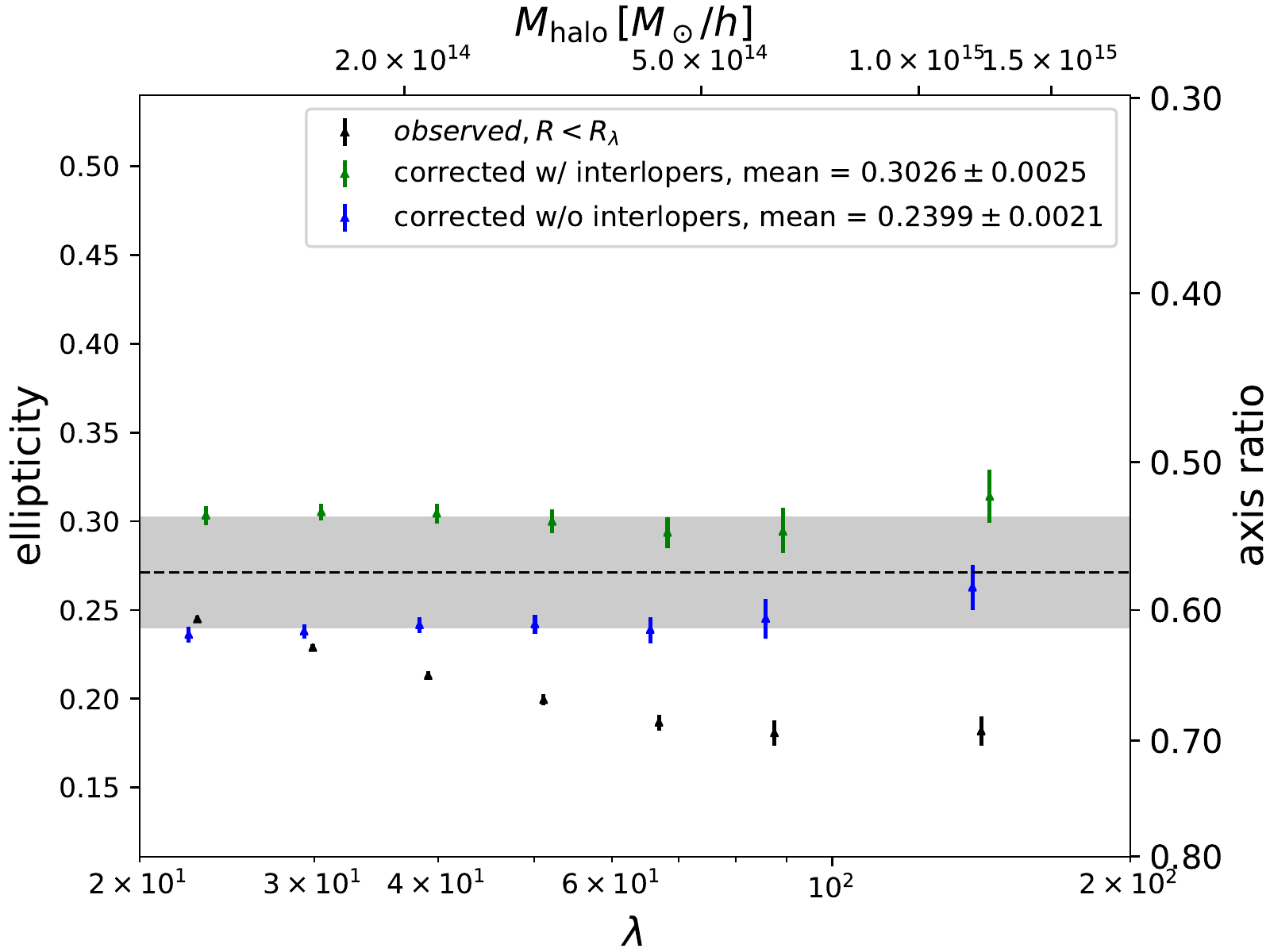}}
\caption{{\it Blue points(A)}: Corrected satellite ellipticities for stacked clusters oriented along the satellite major axis are plotted against cluster richness.
The method of orienting along the satellite axis requires correcting noise and edge bias.
{\it Green points(B)}: The same, but including interlopers in the simulation used for the correction.
Interlopers have a uniform distribution which dilutes the ellipticity. The correction  increases the ellipticity from 0.240 to 0.303.
We take the average of {\it A} and {\it B}  and their difference as our best estimate of the ellipticity and its uncertainty (shown by the grey band): $e = 0.271 \pm 0.031$ corresponding to an axis ratio $q = 0.573 \pm 0.039$. The black points are the uncorrected ellipticity values; the larger correction for richer clusters is due to edge bias. 
}
\label{fig:sim-correct}
\end{figure}

In Fig.~\ref{fig:sim-correct} the black points show the result for the average ellipticity.
The error bars on the observed ellipticity are the standard deviation of the mean within each richness bin.
Looking at these black points, the ellipticity seems to decrease with richness.
However, these raw measurements are subject to several effects that bias the ellipticity of satellite distribution relative to the underlying dark matter halo ellipticity.
In the following sections, we describe the use of simulations to correct for three of these potential biases:
\begin{itemize}
\item ``noise bias'' from measuring ellipticity with a finite number of satellites,
\item ``edge bias'' due to \redmap satellites being restricted to lie within a circular aperture, and
\item bias due to interlopers: red galaxies in the foreground or background that nonetheless make it into the cluster catalog.
\end{itemize}

%%%%%%%%%%%%%%%%%%%%%%%%%%%%%%%%
%%%%%%%%%%%%%%%

%%%%%%%%%%%%%%%
\subsection{Noise and Edge Bias Corrections}
\label{sec:bias}
%%%%%%%%%%%%%%%

Note that edge and noise bias are connected:
noise bias depends on the number of members, and applying a circular edge cut removes some members.
For clarity we first describe the effects of noise bias (Sec.~\ref{sec:noise-bias}) and edge bias (Sec.~\ref{sec:edge-bias}) in isolation.
Then in Sec.~\ref{sec:result-bias} we show the combined effect and the results of correcting both biases in the SDSS data.

%%%%%%%%%%%%%%%
\subsubsection{Noise Bias}
\label{sec:noise-bias}
%%%%%%%%%%%%%%%

The \redmap clusters in our fiducial sample have between 20 and $\sim$ 200 satellites per cluster.
Especially on the low-richness end, the small number of satellites results in a bias towards higher ellipticity values: we call this effect ``noise bias.''
This bias due to measuring ellipticity with a small number of tracers has been noticed in prior work.
For example, \citet{nsd10} noted their Fig.~2 shows that smaller clusters have a larger ellipticity than larger clusters, a result they interpret as coming from this bias.
In addition, \citet{wang08} studied groups with as few as four satellites by using MC simulations to reproduce the full ellipticity distribution to data.
Although their $\chi^2$ values for elliptical halos did not indicate the triaxial model was accurate, these were clearly a much better match to the data than MC isotropic halos.

Like \citet{wang08} we use MC simulations of elliptical halos to deal with this Poisson noise bias.
We use the simulations to correct the observed mean ellipticity.
(Note we ignore the distribution of ellipticity values and assuming each richness bin has just one true ellipticity, but this is checked in Appendix~\ref{sec:ellip-dist}.)
We generate simulated clusters with a range of known ellipticity values, then apply our ellipticity estimator.
The simulations use an elliptical surface density profile,
\be \label{eq:sigma}
\Sigma_{\rm NFW}\left(\sqrt{x^2 + y^2/q^2}\right) \, ,
\ee
\be
\Sigma_{\rm NFW}(r) = \int_{-\infty}^{\infty} \rho_{\rm NFW}(\sqrt{r^2 + \it{z}^2}) \de \it{z} \, ,
\ee
where $\Sigma_{\rm NFW}(r)$ is the 2D NFW profile at projected radius r \citep{nfw1997}, $\rho_{\rm NFW}$ is the 3D NFW profile and {\it z} is line-of-sight distance.
The shape of the profile depends only on mass and concentration.
We use the mass-richness relation of \citet{simet16}.
Concentration is a weak function of cluster mass and therefore richness \citep{msh08}, ranging from $\sim 4.2$ to 5.5 for our clusters \citep{dsk08}.

In our MC simulations, we use the same richness bins as in Fig.~\ref{fig:sim-correct}.
For a given richness bin, we randomly sample richness values from the SDSS \redmap catalog to properly account for fractional abundances of clusters with different richness values.
Then for each such cluster, we treat the normalized surface density of Eq.~(\ref{eq:sigma}) as a probability distribution, and randomly draw $N_{\rm p>0.5}$\footnote{The number of satellites after the $\pmem>0.5$ cut.} sets of $(x,y)$ coordinates from that distribution.
We repeat until the total number of simulated members reaches $\sim 5\times 10^5$, which is more than enough for convergence.

The magnitude of measured ellipticity $e$ is estimated from each cluster following Eqs.~(\ref{eq:e1}) -- (\ref{eq:emag}), then the mean and standard deviation of the mean are calculated.
The results are shown in the left panel of Fig.~\ref{fig:app-nofake} for a range of different richness bins.
Comparing the input (true) and output (measured) values of ellipticity, it is clear that noise bias always increases the observed cluster ellipticity.
The bias is less problematic for larger $\lambda$ values: for fixed true ellipticity, doubling the number of satellites decreases the bias by about a factor of 2.
For all the richness bins, the observed ellipticity asymptotes to the true ellipticity as ellipticity increases. Also, as richness increases, the  measured ellipticity is closer to  the true, input ellipticity.
Note that in the case where we include interlopers (Sec.~\ref{sec:fake-member} and the left panel of Fig.~\ref{fig:app-fake}), the uniform interloper distribution biases the measured ellipticity to a lower value. We discuss the interloper correction in Sec.~\ref{sec:fake-member} in more detail.

Note that correlations between satellites could cause noise bias to be overestimated.
Such correlations have been measured by \citet{fang16}, so we have estimated the magnitude of the effect and confirmed it is negligible.
Details of this estimate are shown in Appendix~\ref{sec:group-infall}.

%%%%%%%%%%%%%%%
\subsubsection{Edge Bias}
\label{sec:edge-bias}
%%%%%%%%%%%%%%%

A second bias results from the requirement that \redmap galaxies fit within a circular aperture of size given by \citet{rykoff14},
\be
R_\lambda = 1 \mpch \, \left(\frac{\lambda}{100}\right)^{0.2} \, .
\ee
This will cut off potential satellites along the major axis, causing an underestimate of the true ellipticity.

As discussed in Sec.~\ref{sec:method}, a choice of $\pmem$ cut can alter the shape of the edge of clusters. If we apply too aggressive a cut on $\pmem$, it will remove almost all the member galaxies on the outskirts. In this case, it becomes a soft-edge cut which depends on the value of $\pmem$ cut instead of the hard-edge cut at $R=R_{\lambda}$ of \redmap's radial filter used to select members. However, the soft-edge is more complicated to model and requires an algorithm-level investigation which is beyond our scope. The $\pmem>0.5$ cut we apply here preserves enough member galaxies around the edge so that the hard edge-cut of \redmap algorithm be still valid.

We can correct for this bias by applying the same $R_\lambda$ cut when measuring the correction factor from simulated clusters.
Applying this same $R_{\rm \lambda}$ cut to the simulations described in Sec.~\ref{sec:noise-bias} leaves $\sim 60\%$ of simulated members. We repeat the same process with this edge cut until the total number of simulated members reaches $\sim 5\times10^5$ per richness bin.
This is enough points for convergence in the mean value and error.

\begin{figure*}
\centering
\resizebox{55mm}{!}{\includegraphics{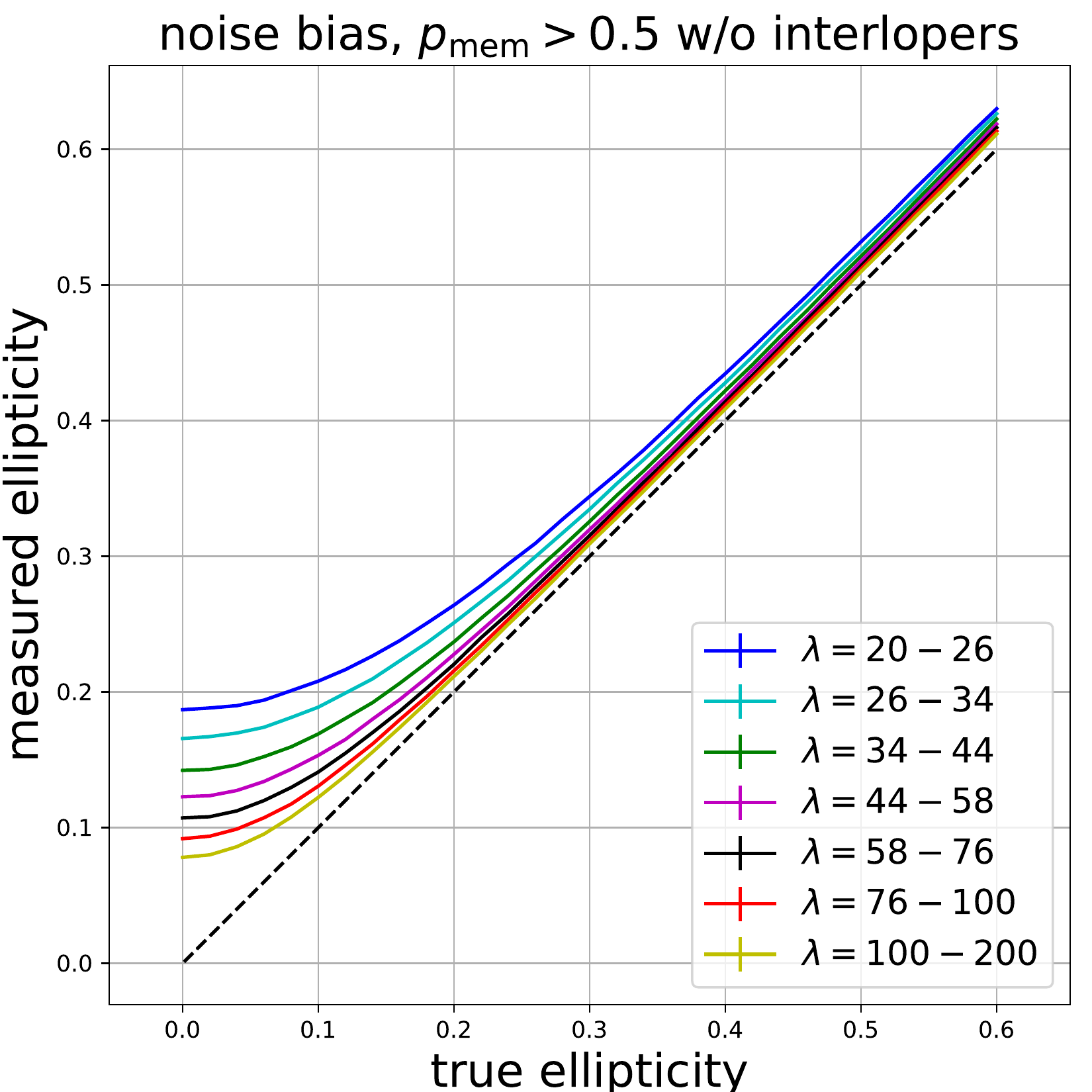}}
\resizebox{55mm}{!}{\includegraphics{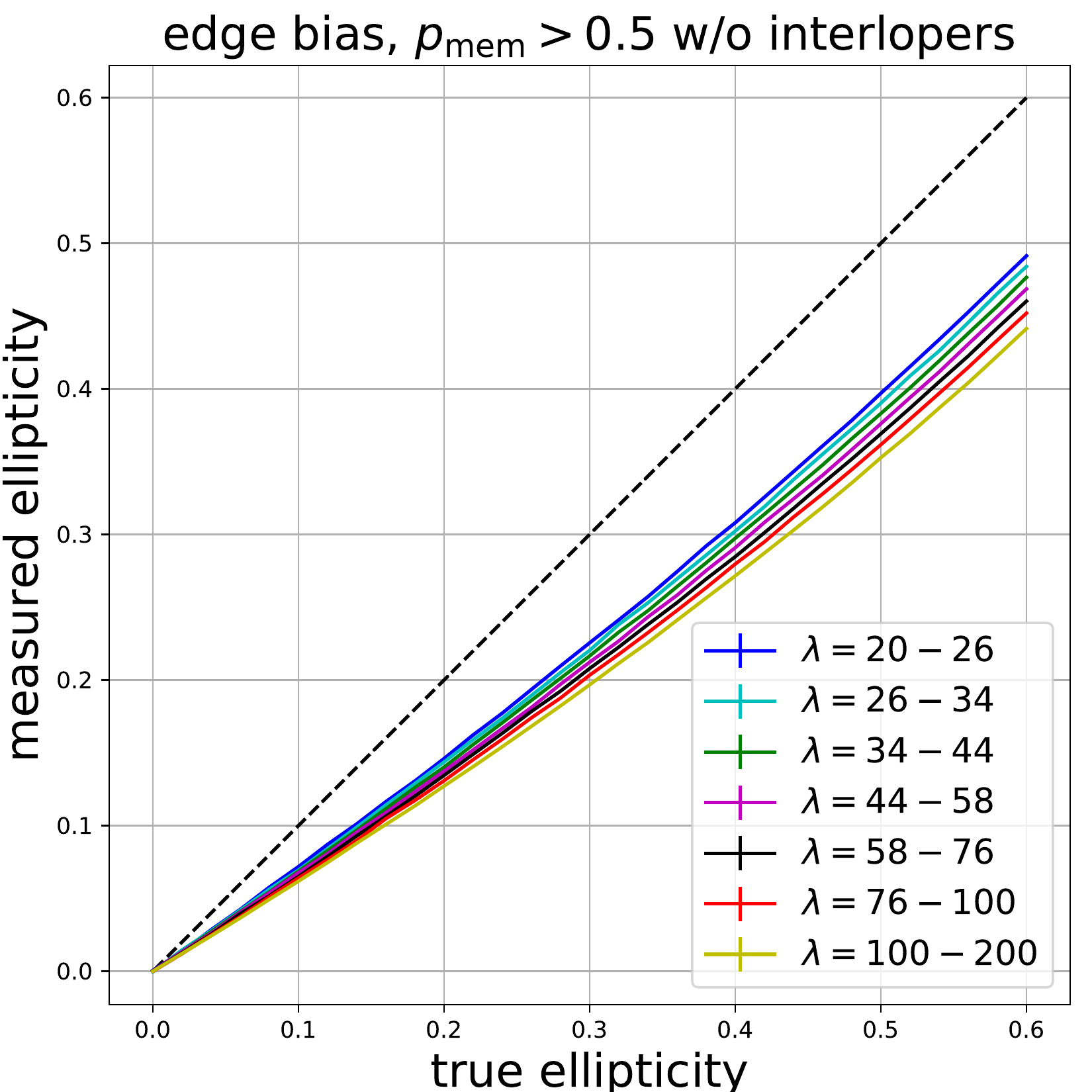}}
\resizebox{55mm}{!}{\includegraphics{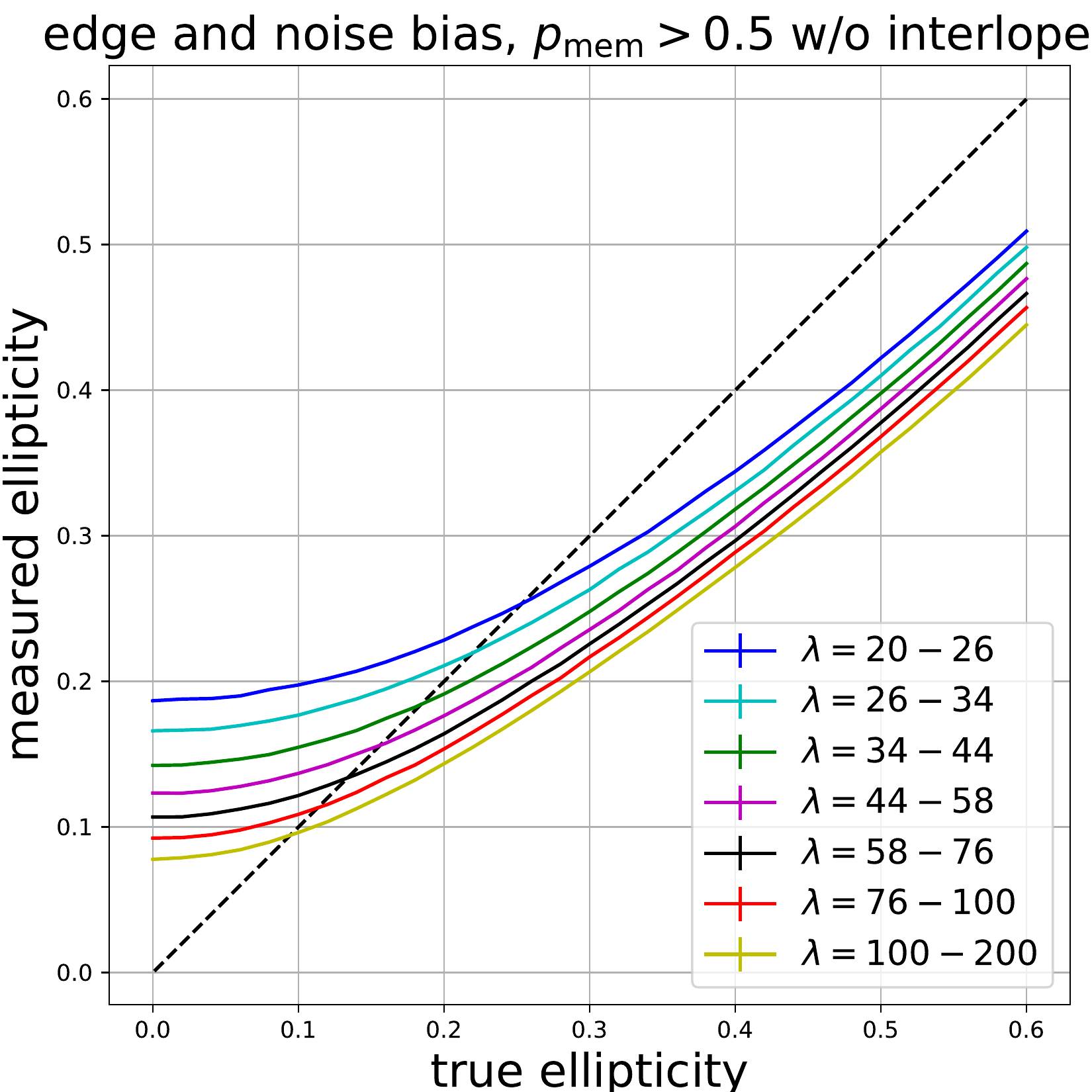}}
\caption{({\it left panel}): Difference between input and measured cluster ellipticity due to Poisson noise bias, for several richness bins.
Noise bias always increases the measured ellipticity.
The effect is strongest for low richness and low ellipticity.
({\it centre panel}): Edge bias due to imposing a circular cut on the elliptical distribution of cluster members.
({\it right panel}): The combined effect of noise and edge bias.
}
\label{fig:app-nofake}
\end{figure*}

\begin{figure*}
\centering
\resizebox{55mm}{!}{\includegraphics{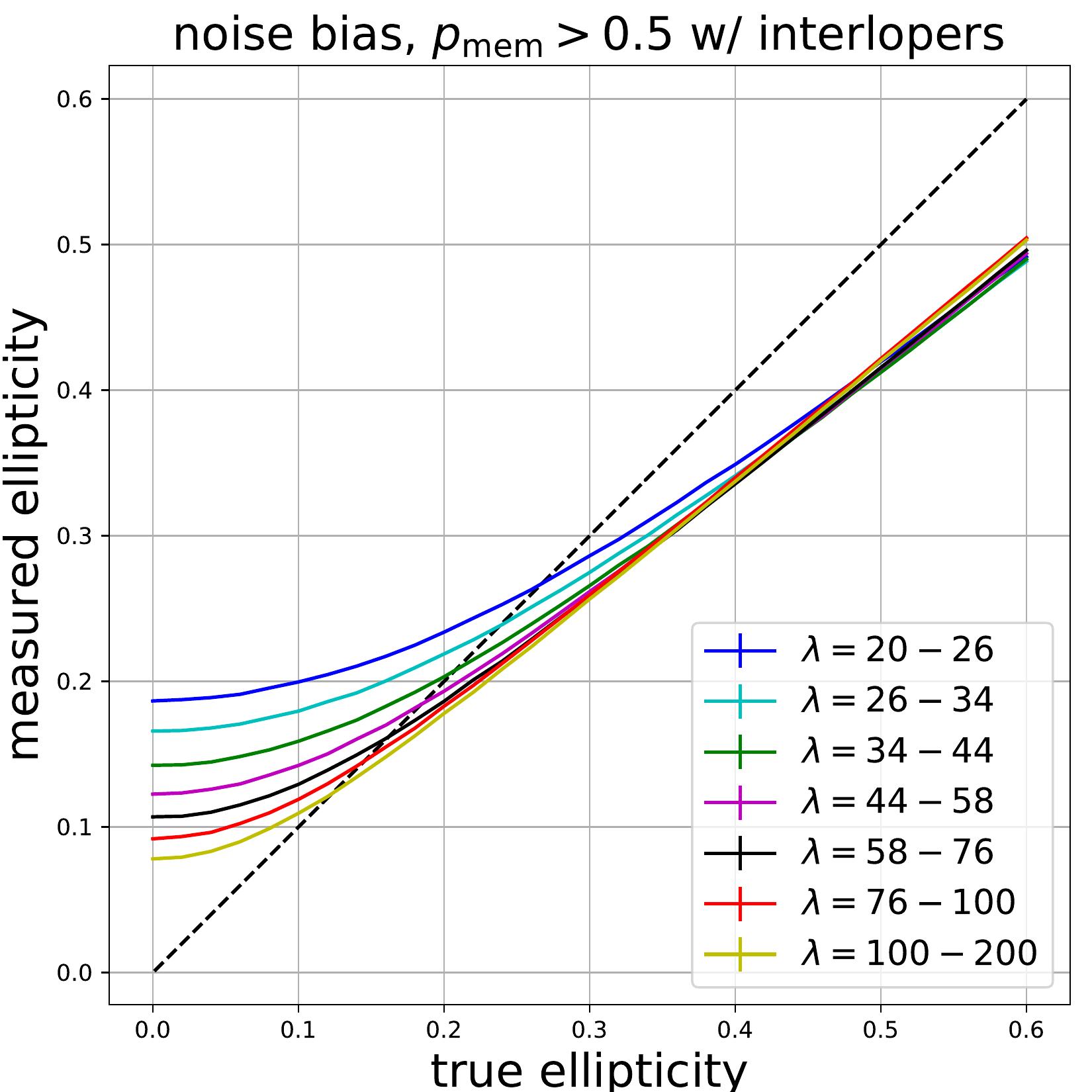}}
\resizebox{55mm}{!}{\includegraphics{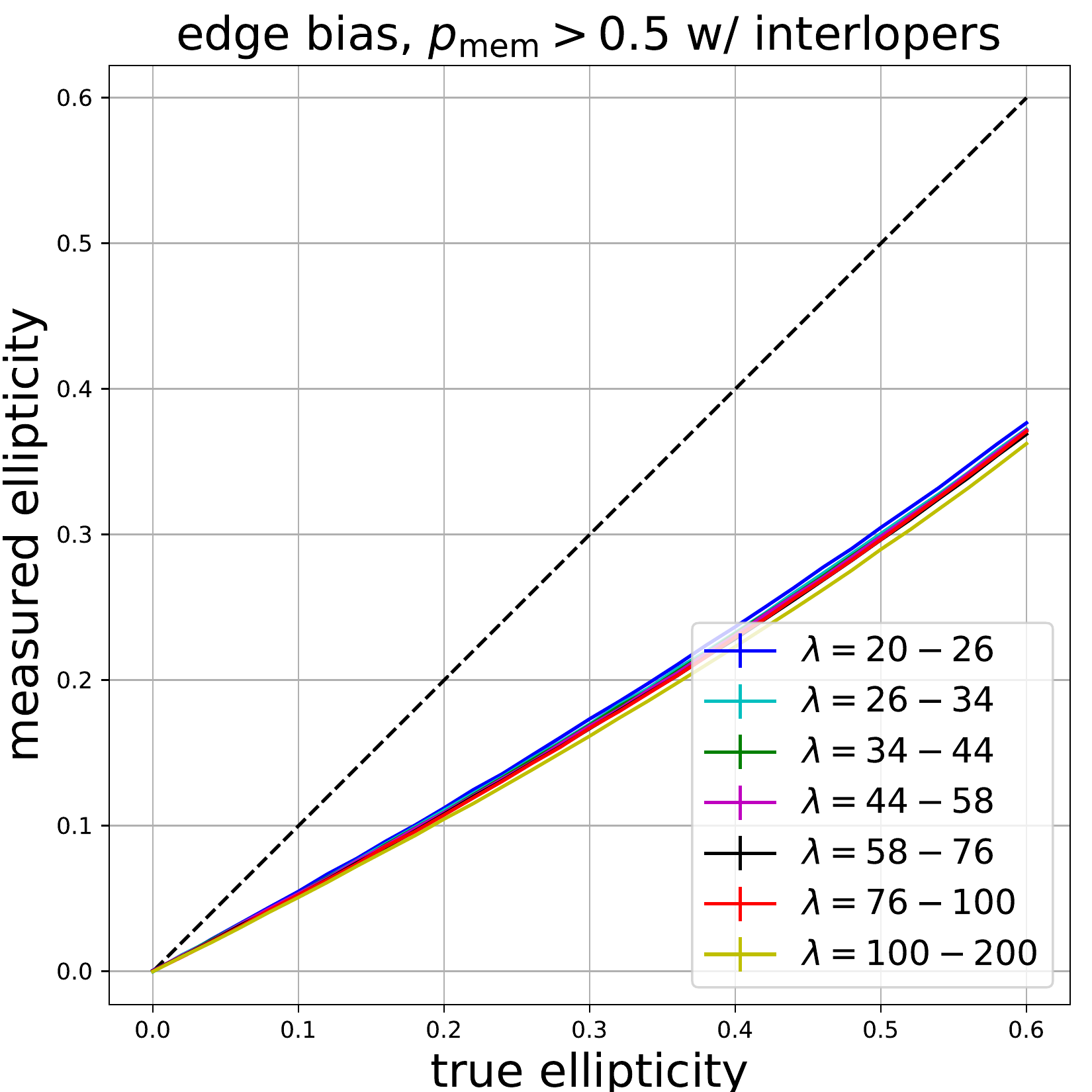}}
\resizebox{55mm}{!}{\includegraphics{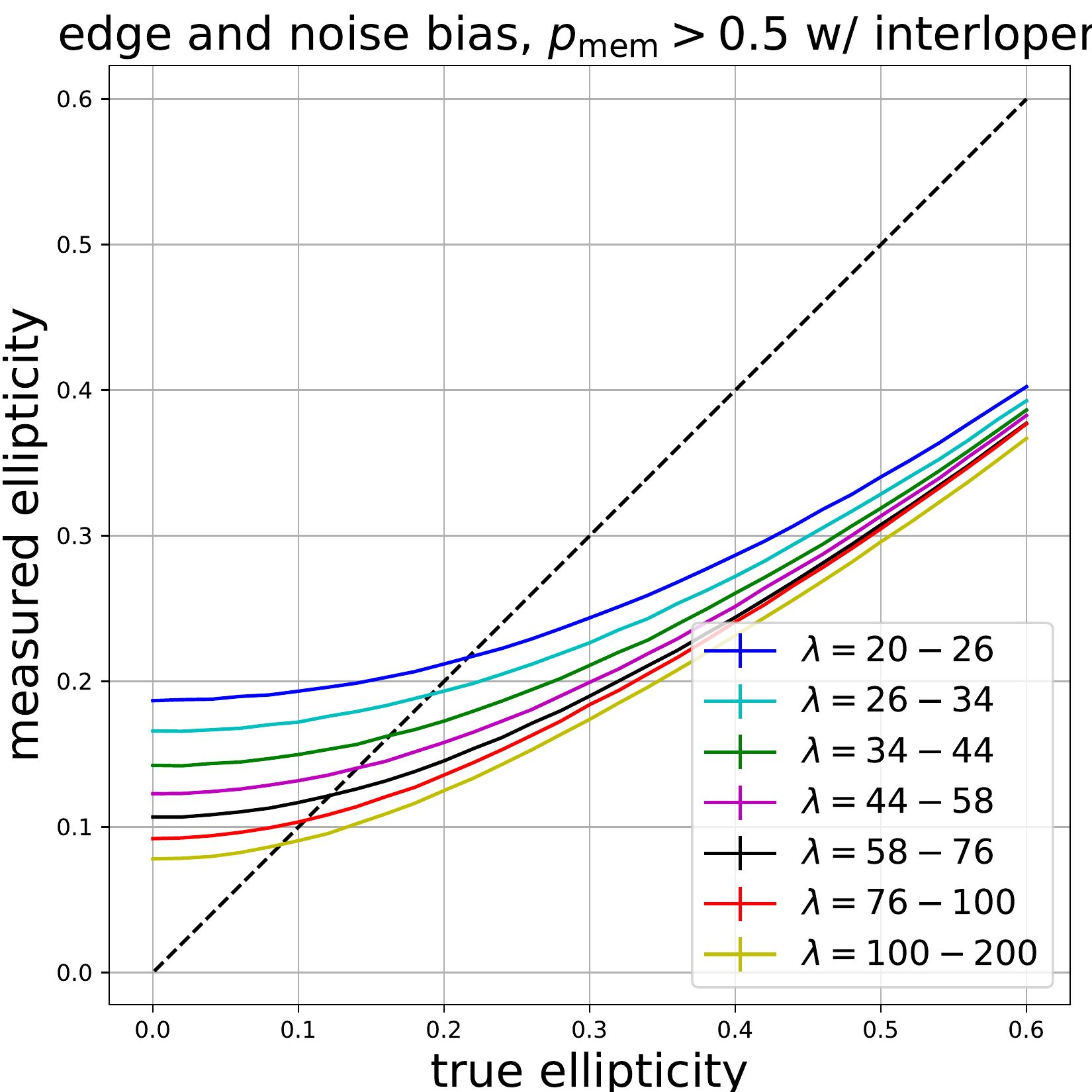}}
\caption{The same as Fig.~\ref{fig:app-nofake}, but including a uniform distribution of fake members or interlopers.
Interlopers result in an underestimate of ellipticity in every case, although for low richness and small input ellipticity, the effect of noise bias still dominates.}
\label{fig:app-fake}
\end{figure*}

For illustration, we first isolate the effect of edge bias by showing only the component of ellipticity ($e_1$) which is aligned with the true major axis of the simulated clusters.
The result is shown in the centre panel of Fig.~\ref{fig:app-nofake}.
As expected, edge bias always tends to decrease the measured ellipticity.
There is no significant dependence on richness.
The fractional error due to edge bias decreases slightly with input ellipticity.

%%%%%%%%%%%%%%%%%%%
%%%%%%%%%%%%%%%%%%%%
%%%%%%%%%%%%%%%%%%%%

%%%%%%%%%%%%%%%
\subsubsection{Results after correcting noise and edge bias}
\label{sec:result-bias}
%%%%%%%%%%%%%%%

In the right panel of Fig.~\ref{fig:app-nofake} we show the combined effect of noise and edge bias in the MC simulations.
Note that edge and noise bias are connected, since noise bias depends on the number of members and applying a circular edge cut removes some members.
In the \redmap data, the richness value indicates the number of members {\it after} the circular edge cut.
So to obtain the corresponding correction from simulations we generate members until the number within the circular aperture $R_\lambda$ reaches $N_{\rm p>0.5}$.

This is the correction we apply to the data, resulting in the blue points in Fig.~\ref{fig:sim-correct}.
The result is well fit by a constant $0.240 \pm 0.002$ (axis ratio 0.613) over an entire decade in richness, from 20 to 200.
This range corresponds to a factor of 20 in halo mass, from $0.88\times10^{14}$ to $1.88\times 10^{15} M_\odot/h$.
These ``corrected'' ellipticity errors are propagated via the mapping between observed and true ellipticity in simulations,
\be
e_{\rm true} \pm \delta e_{\rm true} = f(e_{\rm measured} \pm \delta e_{\rm true})\ 
\ee
where $f$ represents the mapping from measured ellipticity to true ellipticity from the simulation and $\delta e_{\rm true} \, (\delta e_{\rm measured})$ is the uncertainty in $e_{\rm true} \, (e_{\rm measured})$.
Since we assume a constant model for our fit, if there is any relatively strong tendency on ellipticity as a function of richness (see blue dots in Fig.~\ref{fig:sim-correct}), we would underestimate the fitting error by ignoring the non-zero slope of ellipticity as a function of richness.
In this sense, we show the constant fitted line just as an index how elliptical the clusters are in average, but it doesn't mean that clusters in different richness bins would have same ellipticity.

In this section we have studied noise and edge bias using MC simulations.
These biases are straightforward to accurately simulate and correct.
The next section describes a more difficult problem, the correction of biases due to interlopers or ``fake members.''

%%%%%%%%%%%%%%%%%%%%%%%
\subsection{Interloper correction}
\label{sec:fake-member}
%%%%%%%%%%%%%%%%%%%%%

To test the effects of interlopers, we repeat our MC simulations including a population of interlopers along with the cluster members.
For a cluster of richness $\lambda$, we take the members having $\pmem$ values greater than 0.5 and compute $N_{\rm real} = \sum_{i} p_{{\rm mem}, i}$, where the sum is only over those chosen members after $\pmem>0.5$ cut.

As before, these real members are drawn from an elliptical NFW distribution.
Then we add $N_{\rm fake} = \sum_{i} (1- p_{{\rm mem}, i})$ interlopers.
The interlopers are drawn from a uniform circular distribution, i.e., a disk with radius $R_\lambda$.
We use the same richness bins as before, and sample richness and $\pmem$ values from the actual \redmap catalog for consistency between simulation and measurement.

The resulting correction is shown in Fig.~\ref{fig:app-fake}.
Compared to edge and noise bias alone (Fig.~\ref{fig:app-nofake}) adding bias due to interlopers causes an underestimate of the true ellipticity.
We then apply this correction to the raw data (black points) in Fig.~\ref{fig:sim-correct}.
The result (green points) has a higher ellipticity, than the raw data or the previous correction without interlopers (blue points).
This is expected, since the presence of interlopers dilutes the ellipticity further, requiring an even larger correction than in the case with edge and noise bias alone.
The new ellipticity estimate is $0.303 \pm 0.003$, about 26\% higher than the estimate of 0.240 for which interlopers were neglected.

The difference between cases with and without interlopers is $\Delta e = 0.063$.
Since we do not know the actual effect of interlopers, we simply take the average of these extreme cases.
The true value of ellipticity should lie between these two extreme values, assuming no other source of systematics.
The result is an average ellipticity $e = 0.271 \pm 0.031$.
Here we take half the difference as an estimate of the systematic error.

Note that these corrections for the mean ellipticity do not take into account the full distributions of ellipticity within each richness bin, but only the average.
However, after obtaining the best-fit ellipticity values with the simulations, we verified that the simulated ellipticity distributions used for the correction accurately match the real distributions.

%%%%%%%%%%%%%%%
\subsection{Summary}
%%%%%%%%%%%%%%%

We have corrected raw ellipticity measurement of satellites for noise and edge bias.
We also have attempted to correct for interlopers, but this remains our most important systematic.
Our best estimate of the ellipticity of the satellite distribution is 0.271.
Our estimated systematic error is 0.031, from the difference between extreme cases with {\it no interlopers} and $N_{\rm fake}/(N_{\rm fake} + N_{\rm real})$ interloper fraction.
The statistical error is only 0.002 (less than 1\%) assuming constant ellipticity with richness (within richness bins it is in the 2-5\% range). 
Fig. ~\ref{fig:sim-correct} suggests that the corrected ellipticity is indeed approximately constant.

%%%%%%%%%%%%%%%
\section{Weak lensing ellipticity}
\label{sec:lensing}
%%%%%%%%%%%%%%%

%%%%%%%%%%%%%%%
\subsection{Method}
\label{sec:method-lens}
%%%%%%%%%%%%%%%

We use an estimator for halo ellipticity which builds on that developed in \citet{clampitt16}.
We first summarize the model of \citet{clampitt16} before describing more optimal estimators of ellipticity.

This model for the surface density of elliptical halos uses a multipole expansion:
\bea
\Sigma(R, \theta) & \propto & R^{\eta_0} [1 + (-\epsilon \eta_0 (R) /2) \cos{2\theta} + \mathcal{O}(\epsilon^2)] \\
 & \equiv & \Sigma_0 (R) + \Sigma_2 (R) \cos{2\theta} + ...
\eea
and we assume the coeffecient of the quadrupole $-\epsilon \eta_0 / 2 \ll 1$, justifying the neglect of higher orders in the expansion.
Here, $\eta_0$ is logarithmic slope of the monopole, $\de\ln{\Sigma_0}/\de\ln{R}$.
Also, we assume the ellipticity, $\epsilon$, is constant over the whole range of $R$.
Here $R$ is the projected distance from the centre of the halo, and $\theta$ is the angle relative to the halo's major axis.
We set
\be \label{eq:assume}
\epsilon \approx -\frac{2 \Sigma_2 (R)}{\eta_0 (R) \Sigma_0 (R)} \, ,
\ee
thus allowing the quadrupole $\Sigma_2$ to be completely determined by the monopole $\Sigma_0$, up to a proportionality factor $\epsilon$, the magnitude of the ellipticity.
In order to avoid confusion with various ellipticity definitions, we actually fit the more universal axis ratio, $q$, given by $\epsilon = (1 - q^2) / (1+q^2)$.

Recall that the Cartesian components of shear can be derived from the tangential and cross components as
\bea
\gamma_1 & = & -\gamma_+ \cos{2\theta} + \gamma_\times \sin{2\theta} \\
\gamma_2 & = & -\gamma_+ \sin{2\theta} - \gamma_\times \cos{2\theta} \, .
\eea
\citet{clampitt16} builds on \citet{acd15} to derive the following equations for the quadrupole shear in Cartesian coordinates:
\bea
\Sigma_{\rm crit} \gamma_1 (R,\theta) & = & (\epsilon/4) \, [(2 I_1(R) -\Sigma_0(R) \eta_0(R)) \cos{4\theta} + \nonumber \\
& & 2 I_2(R) - \Sigma_0 (R) \eta_0 (R)] \, , \label{eq:gamma1} \\
\Sigma_{\rm crit} \gamma_2 (R,\theta) & = & (\epsilon/4) \, [2 I_1 (R) - \Sigma_0 (R) \eta_0(R)] \sin{4\theta} \, . 
\eea
where
\bea
I_1 (R) & \equiv & \frac{3}{R^4} \int_0^R R'^3 \Sigma_0 (R') \eta_0(R') \de R' \, , \\
I_2 (R) & \equiv & \int_R^\infty \frac{\Sigma_0 (R')}{R'} \eta_0(R') \de R' \, ,
\eea
\bea \label{eq:crit}
\Sigma_{\rm crit} & \equiv & \frac{c^2}{4\pi G} \frac{D_{A} (z_{s})}{D_{A} (z_{l}) D_{A} (z_l, z_s)} \, ,
\eea
where $D_{A}(z_{s}), D_{A}(z_{l})$ and $D_{A}(z_{l}, z_{s})$ represent angular diameter distance from observer to source, from observer to lens and from lens to source, respectively.
We go a step further than \citet{clampitt16} and define optimally-weighted\footnote{By ``optimal'', we mean that the full (internal plus external) quadrupole signal is captured with a weight that maximizes S/N and further that we have separated the internal and external quadrupole signal.} halo ellipticity estimators that include all the information from both Cartesian components:
\bea
\Delta\Sigma^{4\theta} & \equiv & \frac{1}{2} \frac{1}{2\pi} \int \de\theta \left(\frac{\Sigma_{\rm crit} \gamma_1 (R,\theta)}{\cos{4\theta}} + \frac{\Sigma_{\rm crit} \gamma_2 (R,\theta)}{\sin{4\theta}}\right) \label{eq:4th_theory} \\
 & = & (\epsilon / 4) [2 I_1(R) - \Sigma_0(R) \eta_0(R)] \label{eq:4th_theory2}
\eea
and
\bea
\Delta\Sigma^{\rm const} & \equiv & \frac{1}{2\pi} \int \de\theta \, \Sigma_{\rm crit} \gamma_1 (R,\theta) \label{eq:const_theory} \\
 & = & (\epsilon / 4) [2 I_2(R) - \Sigma_0(R) \eta_0(R)] \, ,  \label{eq:const_theory2}
\eea
which \citet{bernstein09} show come from internal and external quadrupoles, respectively.
This useful distinction between quadrupoles internal and external to $R$ will be described in greater detail when we present our results in Sec.~\ref{sec:results}.

In the presence of misalignment between the light and dark matter major axes the measured ellipticity is actually an underestimate of the halo ellipticity:
\be
\epsilon^{\rm observed} = D \times \epsilon^{\rm true} \, .
\label{eq:D}
\ee
The misalignment factor is given by
\be \label{eq:dilute}
D \equiv \int \, \de \theta_{\rm off} \, P\,(\theta_{\rm off}) \, \cos{(2\theta_{\rm off})} \, ,
\ee
where $P\,(\theta_{\rm off})$ is the distribution of angle differences $\theta_{\rm off}$ between the major axes of dark matter and the visible proxy (measured major axis of each cluster's satellite distribution according to Sec.~\ref{sec:method}).
See \citet{clampitt16} for the derivation of this equation. 

For clarity, we note that we have used three different definitions of major axis in this paper. These are: 
\begin{itemize}
\item Major axis of the underlying distribution according to which member galaxies are scattered (A),
\item Major axis of distribution of member galaxies for a given cluster, calculated as in Sec.~\ref{sec:method} (B) 
\item Major axis of dark matter halo (C).
\end{itemize}
B is subject to a Poisson noise bias with respect to A (see Sec.~\ref{sec:noise-bias}).
Assuming A and C are perfectly aligned, we calculate the misalignment between B and C.
For the latter in Appendix~\ref{sec:misalign} we give the relation between misalignment and richness, $D(\lambda)$, as well as misalignment distributions for several values of ellipticity and $\lambda$.
The MC simulations described in Sec.~\ref{sec:satellite} were used to obtain these relationships.

Since we have 10428 different clusters in the catalog, the expected signal would be the average of those clusters.
Using Eqs.~(\ref{eq:4th_theory}-\ref{eq:const_theory2}) and the misalignment factor described above, we calculate the expected average signal for a range of ellipticity to be fitted to our measurements, adopting the misalignment factors calculated with the ellipticity of the satellite distribution.
We discuss our model in more detail in Sec.~\ref{sec:model}.

%%%%%%%%%%%%%%%
\begin{table*}
\centering
Comparison of axis ratios\\
\begin{tabular}{l|c|c|c|c|c}
\hline
Stacking Method & Number of clusters & Corrected for Poisson sampling & Satellite axis ratio & Lensing axis ratio & Lensing significance \\
\hline
Satellites & 10428 & yes & $0.573 \pm 0.039$ & $0.558 \pm 0.086 \pm 0.026{\rm (sys)} $ & 5.1 $\sigma$ \\
Satellites & 10428 & no & & $0.649 \pm 0.067$ & 5.2 $\sigma$ \\
Central Galaxy & 6681 & not applicable & $0.772 \pm 0.025$ & $0.746 \pm 0.058$ & 4.4 $\sigma$ \\
\end{tabular}
\caption{Comparison of axis ratios from lensing and the satellite distribution.
The lensing axis ratio includes statistical error bars mostly dominated by shape noise.
The satellite axis ratio error is dominated by systematic uncertainty in the number of interlopers.
The first row shows the axis ratio corrected for the misalignment between satellites and individual halos due to Poisson sampling of the satellites.
Agreement between visible galaxies and lensing is very good for both stacking methods.
}
\end{table*}
%%%%%%%%%%%%%%%

%%%%%%%%%%%%%%%
\begin{figure}
\centering
\resizebox{85mm}{!}{\includegraphics{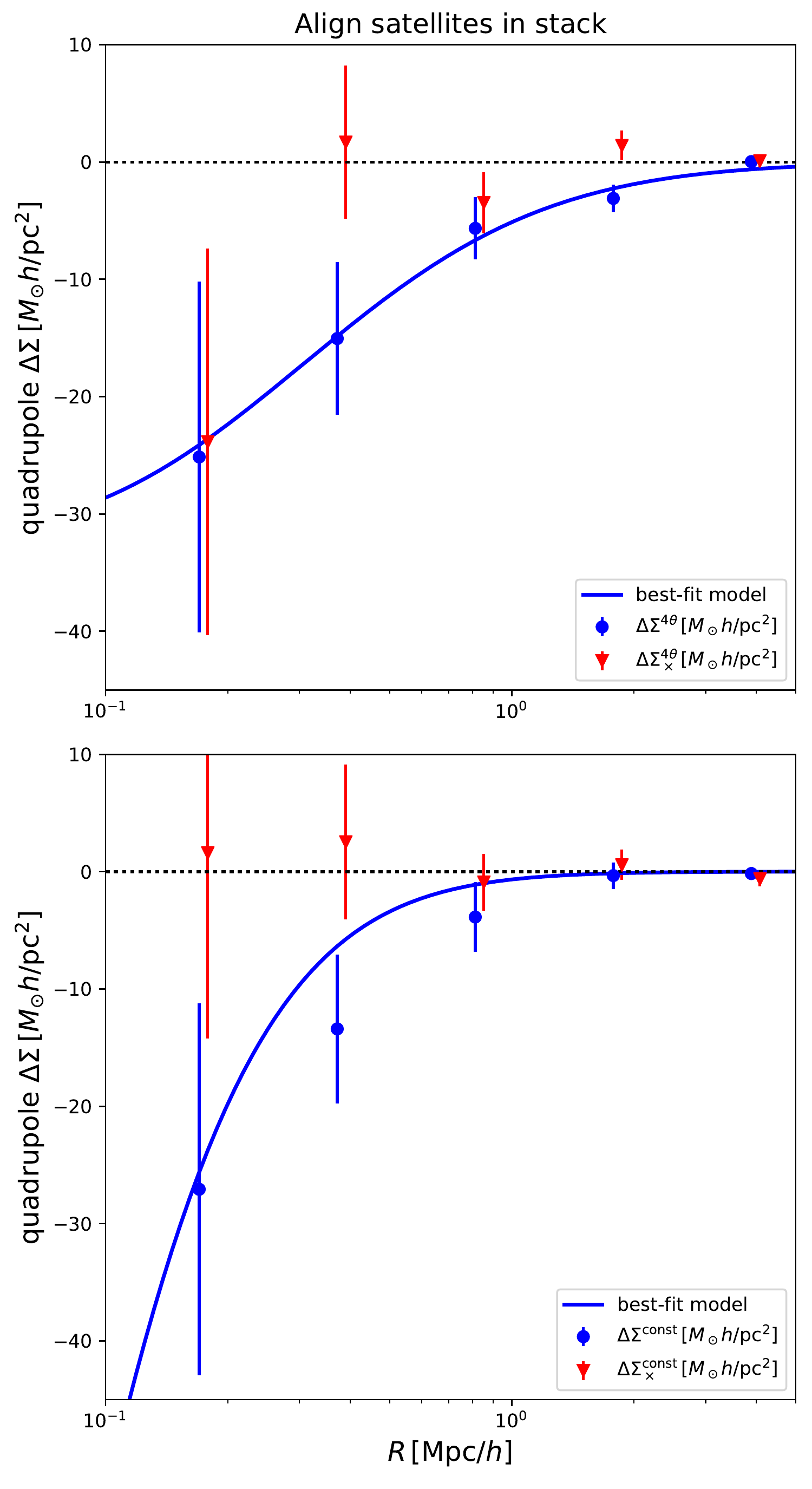}} \\
\caption{Weak lensing quadrupole signal of  clusters (blue points), stacked on the major axis of the satellite distribution.
The best-fit model (blue line) and lensing cross-components (red triangles) are also shown.
The two components of the quadrupole signal, $\dstheta$ (top panel) and $\dsconst$ (bottom panel), are subject to different systematics.
The signal indicates a best-fit axis ratio of $q = 0.558 \pm 0.086$, corresponding to a $\sim 5\sigma$ detection, while the cross-components are consistent with zero.
}
\label{fig:lensing-sig-sat}
\end{figure}
%%%%%%%%%%%%%%%

%%%%%%%%%%%%%%%
\subsection{Ellipticity estimator}
\label{sec:estimator}
%%%%%%%%%%%%%%%

Equation~(\ref{eq:4th_theory}) implies the optimal estimator for the varying component of the quadrupole lensing signal is
\be
\langle \Delta\Sigma^{4\theta} \rangle (R_j) = B(R_j) \frac{\sum_i \Sigma_{{\rm crit},ij} \left(\frac{w_{1,ij} \gamma_{1,ij}}{\cos{4\theta_{ij}}} + \frac{w_{2,ij} \gamma_{2,ij}}{\sin{4\theta_{ij}}} \right)}{\sum_i w_{1,ij} + w_{2,ij}} \, ,
\ee
where
\bea
w_{1,ij} = \cos^2{\!(4\theta_{ij})} \; \Sigma^{-2}_{{\rm crit},ij} / (\sigma_{\rm shape}^2 + \sigma_{{\rm meas},ij}^2) \\
w_{2,ij} = \sin^2{\!(4\theta_{ij})} \; \Sigma^{-2}_{{\rm crit},ij} / (\sigma_{\rm shape}^2 + \sigma_{{\rm meas},ij}^2) \, ,
\eea
where i runs over each lens-source pair inside each j-th radial bin, $R_j$ and $\Sigma_{{\rm crit},ij}$ follows the definition of Eq.~(\ref{eq:crit}). $B(R)$ is the boost factor accounting for the correlated galaxies that are accidentally included in the lens (see the following paragraph for the definition and details). We have suppressed the $R_j$ arguments of the shear components for simplicity.
The noise terms of the shear estimator $\gamma$ are the usual shape noise due to intrinsic galaxy ellipticities, $\sigma_{\rm shape}=0.32$, and the shape measurement error, $\sigma_{{\rm meas},ij} \sim 0.05$.
Likewise, Eq.~(\ref{eq:const_theory}) implies the optimal estimator for the constant (with angle) quadrupole lensing signal is
\be
\langle \Delta\Sigma^{\rm const} \rangle (R_j) = B(R_j) \frac{\sum_i (w_{ij} \Sigma_{{\rm crit},ij} \gamma_{1,ij})}{\sum_i w_{ij}} \, ,
\ee
where
\be
w_{ij}= \Sigma^{-2}_{{\rm crit},ij} / (\sigma_{\rm shape}^2 + \sigma_{{\rm meas},ij}^2) \, .
\ee
note that $w_{ij}$ here is different from $w_i$ in Sec.~\ref{sec:method}.

The boost factor $B(R) > 1$ is used to correct the dilution of signal at small scales from sources that are actually physically correlated with the lens.
We make use of the random point catalog accompanied by the \redmap cluster catalog in Sec.~\ref{sec:rm-data}.
Then the boost factor is calculated according to (\citealt{sjf2004, msh08})
\be
B(R_j) = \frac{N_{{\rm rand},j}}{N_{{\rm lens},j}} \frac{\sum_i w_{ij}}{\sum_k w_{kj}} \, ,
\ee
where $i$ indicates lens-source pairs, $k$ indicates random-source pairs and $N_{{\rm lens},j}$ ($N_{{\rm rand},j}$) is the number of lens clusters (random points) inside the j-th radial bin.
Note that we see no evidence for azimuthal variation in the boost factor, in agreement with the results of \citet{van16}.
The boost factors for satellite stacking and CG major axis stacking is the same, since the richness and redshift distributions of both samples are nearly identical.

The lensing cross-components are given by
\be
\langle \Delta\Sigma^{4\theta}_\times \rangle (R_j) = B(R_j) \frac{\sum_i \Sigma_{{\rm crit},ij} \left(\frac{w_{1,ij} \gamma_{1,ij}}{\cos{4\theta_{ij}}} - \frac{w_{2,ij} \gamma_{2,ij}}{\sin{4\theta_{ij}}} \right)}{\sum_i w_{1,ij} + w_{2,ij}} \, ,
\ee
and
\be
\langle \Delta\Sigma^{\rm const}_\times \rangle (R_j) = B(R_j) \frac{\sum_i (w_{ij} \Sigma_{{\rm crit},ij} \gamma_{2,ij})}{\sum_i w_{ij}} \, .
\ee
Note that the quadrupole cross-components do not vanish identically for all lensing like they do for monopole.
They only vanish if the mass distribution is symmetric under reflections about the $x$- and $y$-axes.
Fig.~\ref{fig:sat-contour} shows the contours of the satellite distribution are symmetric under reflections.
Furthermore, since we are stacking $\sim$ 10,000 clusters (with major axes aligned along the $x$-axis) it is a safe assumption that the mass contours are also symmetric after stacking.
The asymmetry of individual cluster halos will be smoothed out in the stack so we expect these cross-components to qualify as null tests.

Note that we have not weighted each lens-source pair by lens ellipticity.
It is difficult to obtain a reliable ellipticity for individual cluster halos, due to the biases described in Sec.~\ref{sec:bias}.
And when stacking using the CG light distribution (as in Sec.~\ref{sec:cg-stack}) we see no evidence that weighting by lens ellipticity improves the signal-to-noise ratio (S/N).
A similar test was carried out by \citet{van16} in their study of group ellipticities from weak lensing.
In Section 4.1 of \citet{van16} find slightly weaker constraints when weighting by CG ellipticity.
They claim this is not surprising since their KSB shape measurement upweights the bulge at the centre of the light.
The bulge is rounder and also may not correlate as well as the outer part of the light profile with the dark matter.

For all lensing measurements we obtain the covariance using 200 jackknife patches \citep{nbg09} generated from a \texttt{kmeans} algorithm\footnote{\texttt{https://github.com/esheldon/kmeans\char`_radec}}.
This algorithm divides the lens clusters into 200 approximately equal area patches over the entire SDSS survey.
We use the full resulting covariance matrix for all fits and quoted S/N values.

%%%%%%%%%%%%%%%
\begin{figure}
\centering
\resizebox{85mm}{!}{\includegraphics{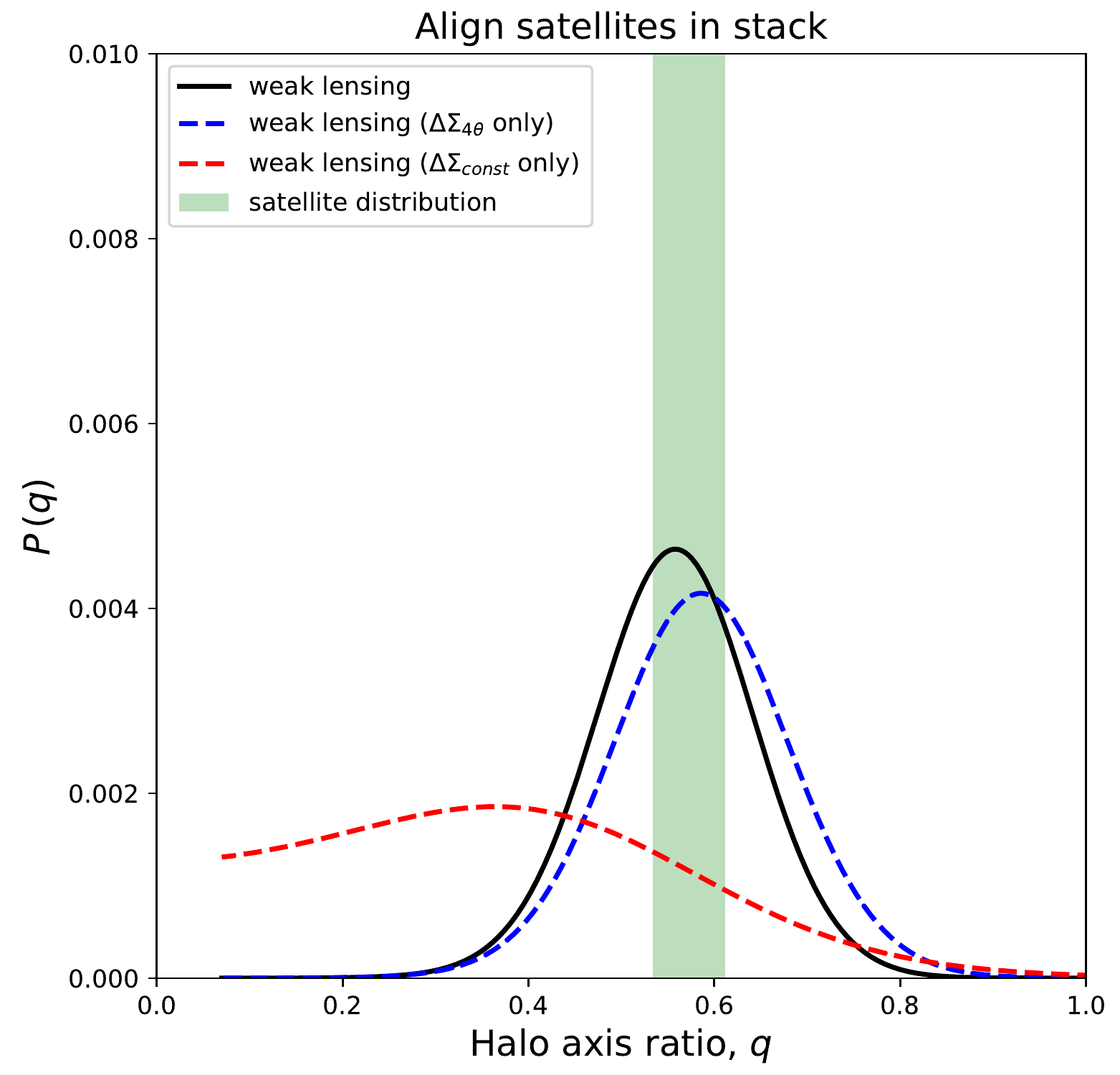}}
\caption{The posterior, $P(q)$, of the axis ratio, $q$, of the cluster halos when orienting the stacked weak lensing measurement along the major axis of each cluster's satellite distribution.
The results with both signal components ($\dstheta$ and $\dsconst$, black solid line) and only $\dstheta$ (blue dashed line) are consistent.
The $\dsconst$-only posterior (red dashed line) is very loosely constrained.
In addition, the axis ratio of the satellite distribution (green shaded band) is consistent with the lensing results.
Note that both satellite and lensing axis ratios have been corrected for the effect of Poisson sampling: this is necessary to make a fair comparison (see the text for details).
}
\label{fig:lensing-fit-sat}
\end{figure}
%%%%%%%%%%%%%%%

%%%%%%%%%%%%%%%%%
\subsection{Model}
\label{sec:model}
%%%%%%%%%%%%%

Next we describe the model in more detail.
To fit the halo ellipticity we use an NFW model for the monopole and, for a given true ellipticity $\epsilon$, we calculate the quadrupole  according to Eqs.~(\ref{eq:4th_theory2}) and (\ref{eq:const_theory2}).
Since an NFW profile depends on the mass of the halo, we need information on the mass of the clusters. 
Instead of using one effective mass representing the whole cluster sample, we calculate a more accurate  model as follows.

For the mass-richness calibration of this cluster sample, we use the results of \citet{simet16},
\be \label{eq:mass-rich}
M_{200} = 2.21\times 10^{14} M_\odot/h \times \left( \frac{\lambda}{40}\right)^{1.33} \, .
\ee
There are uncertainties in the parameters of the mass-richness relation. However, we only use the mean values of the parameters as shown in the above equation when calculating the mass of each cluster. We have tested that our result is insensitive to the scatter in the mass-richness relation.
The corresponding NFW profiles are the monopoles of the clusters.
Then using Eqs.~(\ref{eq:4th_theory2}) and (\ref{eq:const_theory2}) we obtain the quadrupole for internal and external contributions, $\Delta\Sigma^{4\theta}$ and $\Delta\Sigma^{const}$ respectively.
Finally, we apply misalignment factor D (Sec.~\ref{sec:method-lens} and below) to each cluster and average these model quadrupole signals ($D\Delta\Sigma^{4\theta}$ \& D$\Delta\Sigma^{const}$) to get the final  quadrupole.
We generate models with ellipticity varying from 0 to 0.9, assuming the ellipticity is constant with radius. 
Using these models, we fit the axis ratio instead of ellipticity to the measured quadrupole signal and obtain the final constraint on axis ratio.

The misalignment factor D is calculated as follows. 
We estimate in Sec.~\ref{sec:satellite} that the true underlying ellipticity of satellite distribution is 0.271.
Using the same MC simulation as Sec.~\ref{sec:satellite} assuming $e_{\rm true}=0.271$, we can calculate the misalignment factor D as a function of number of satellite galaxies used in major axis calculation ($N_{sat}$).
With an assumption that $D(N_{sat}) = D(\lambda)$ we  then obtain the misalignment factor $D$ to be applied to every individual cluster. We plot $D(N_{sat})$ in Fig.~\ref{fig:dilution}.
We propagate the systematic uncertainty $\Delta e_{\rm true} = 0.031$ into the model through the calculation of $D$.

Note that our model does not include the 2-halo term which can alter the outer profiles, especially as we approach the maximum radius we measure,  $5 \mpch$. If the 2-halo term impacts the monopole and quadrupole differently than the 1-halo term, it would impact the ellipticity.\footnote{The 2-halo term refers to the effect of the surrounding matter around the mass profile of the halo of interest.}

%%%%%%%%%%%%%%%
\subsection{Results}
\label{sec:results}
%%%%%%%%%%%%%%%

The quadrupole lensing signal is shown in Fig.~\ref{fig:lensing-sig-sat}.
The measurements of both components $\dstheta$ and $\dsconst$ are significant over scales $0.1 \lesssim R \lesssim 2 \mpch$.
Both components also have the right sign (negative) for lensing from halo ellipticity.
The detection significance, 
\be
\frac{1 - \text{best-fit axis ratio}}{\text{uncertainty of best-fit axis ratio}} \, ,
\ee
is $5 \sigma$, with more signal coming from the $\dstheta$ component.
The corresponding cross-components $\dsthetax$ and $\dsconstx$ are both consistent with zero, with $\chi^2$ per degree of freedom being 1.03 and 0.29 respectively.
We have also checked that the lensing ellipticity constraints are not very sensitive to the centering probability: using 80\% or 95\% (instead of our fiducial value of 90\%) shifts the best-fit value by well under $1\sigma$.

As expected from Eqs.(\ref{eq:4th_theory2}) and (\ref{eq:const_theory2}) the constant term falls off faster than the $4\theta$ term.
The $\dsconst$ signal is consistent with zero at $\sim 1 \mpch$ while the $\dstheta$ signal continues out to $\sim 3 \mpch$.
Physically, this happens because of the nonlocal nature of both components of shear.
Recall that for the monopole signal $\Delta\Sigma (R)$ or $\gamma_+$ only mass inside $R$ contributes to the signal.
Circularly-symmetric (i.e., monopole) mass external to $R$ causes no lensing when averaged over the entire annulus.
In contrast, mass arranged in a quadrupole produces two distinct signal components \citep{bernstein09} depending on whether the quadrupole mass is {\it internal} ($\dstheta$) or {\it external} ($\dsconst$) to $R$.
Thus, if the shear is produced by an elliptical halo confined to $R \lesssim 1 \mpch$, we would expect $\dstheta$ to fall off rapidly beyond $R \sim 1 \mpch$ since at that point none of the halo is external to $R$. The measurements in Fig.~\ref{fig:lensing-sig-sat} match the predictions for an NFW halo out to $3 \mpch$. 

 On the other hand, $\dsconst$ is sensitive to quadrupoles external to the halo as well, 
 perhaps produced by filaments or large-scale structure.
In Fig.~\ref{fig:lensing-sig-sat} we also show the best-fit model of the 1-halo term quadrupole.
At intermediate scales $\sim 0.5 \mpch$ the $\dsconst$ signal is slightly stronger than the best-fit model. 
This excess could include contributions  from filaments or other external quadrupoles.
Higher S/N data would be needed to conclusively show whether or not our measurement has a contribution from external quadrupoles.
However, we can get some intuition for the contribution from Fig.~\ref{fig:lensing-fit-sat} which compares the significance of the joint fit ($5\sigma$) to the significance of $\dstheta$ alone ($4.3\sigma$).
The two fits are not in tension.

When fitting the axis-ratio we calculate the likelihood $\mathcal{L}$ as 
\be
{\rm ln}\mathcal{L} = -\frac{1}{2} \sum_{ij} (d_{\rm d} - d_{\rm m})_i \,Cov^{-1}_{ij}\, (d_{\rm d} - d_{\rm m})_j \, ,
\ee
where $d_{\rm d}$ and $d_{\rm m}$ are vectors of measured data points and model expectation respectively and $Cov^{-1}_{ij}$ is the (i, j) component of the inverse covariance matrix. Note that  we combine $\Delta\Sigma^{4\theta}$ and $\Delta\Sigma^{const}$ signal when calculating $\mathcal{L}$.
Then we apply the maximum-likelihood method to get the best-fit axis ratio q and its uncertainty, with a non-informative prior on q over [0,1.2].
The full posterior for the axis-ratio fit is shown in Fig.~\ref{fig:lensing-fit-sat}.
The best-fit value is $q = 0.558 \pm 0.086$(stat) $\pm 0.026$(sys), or $e = 0.284 \pm 0.072$(stat) $\pm 0.021$(sys), where the statistical error is dominated by lensing shape noise\footnote{Jackknife error estimation reflects both shape noise and cluster-to-cluster variance. To confirm that the statistical error is dominated by shape noise, we perform an additional test by randomly rotating the source shapes to wash out cluster-to-cluster variance. The errors on $\dstheta$ and $\dsconst$ then change by $\sim 5\%$ at most, which indicates our error is dominated by shape noise.}and the systematic uncertainty originates from the uncertainty in ellipticity of satellite distribution, which is used to calculate the misalignment factor $D$.
This is in an agreement within $1 \sigma$ with the axis ratio of the satellite distribution, $0.58$, whose 68\% confidence interval is also shown on Fig.~\ref{fig:lensing-fit-sat} for comparison.
Note that for the satellite result, we converted the error on ellipticity (0.031) estimated in Sec.~\ref{sec:satellite} to an error on the axis ratio (0.039) via Eq.~(\ref{eq:axis}).
Note that if we do not correct for Poisson sampling effect of satellites (the misalignment factor D in Eq.~\ref{eq:D}), the measured best fit axis ratio from lensing 
is $q=0.649 \pm 0.067$, or $e = 0.213 \pm 0.049$.
Comparing the two results with and without Poisson sampling correction, we deduce $D = 0.213/0.284 = 0.750$ with an uncertainty of 0.059, where the uncertainty is evaluated using covariances from the same jackknife patches as in Sec~\ref{sec:estimator}. This translates into an RMS of $18^\circ \pm 2.5^\circ$ between the true major axes and the major axes of the observed satellite distribution (the stacking orientation).

We also fit the axis ratio to $\dstheta$ and $\dsconst$ separately as a consistency test. The result is shown by the blue and the red dashed curves respectively in  Fig.~\ref{fig:lensing-fit-sat}. The best fit axis ratios of $\dstheta$ and $\dsconst$ are $q = 0.585 \pm 0.097$(stat) $\pm 0.024$(sys) and $q=0.365 \pm 0.207$(stat) $\pm 0.044$(sys), consistent with our fiducial result.

Since both lensing and satellite axis ratios have been corrected for Poisson sampling, 
the posteriors in Fig.~\ref{fig:lensing-fit-sat} represent our constraints on the true satellite distribution and the {\it dark matter} halo axis ratios, respectively, of \mbox{\redmap} clusters.
The agreement of the lensing with the satellite distribution is striking and rules out any significant misalignment between the satellite galaxies and the cluster halo.
Likewise, the agreement between our measurements and CDM N-body simulations is excellent.
In Sec.~\ref{sec:compare} we describe those comparisons in more detail.
In Table 1 we summarize our main results for lensing and satellite axis ratios.

%%%%%%%%%%%%%%%
\begin{figure}
\centering
\resizebox{87mm}{!}{\includegraphics{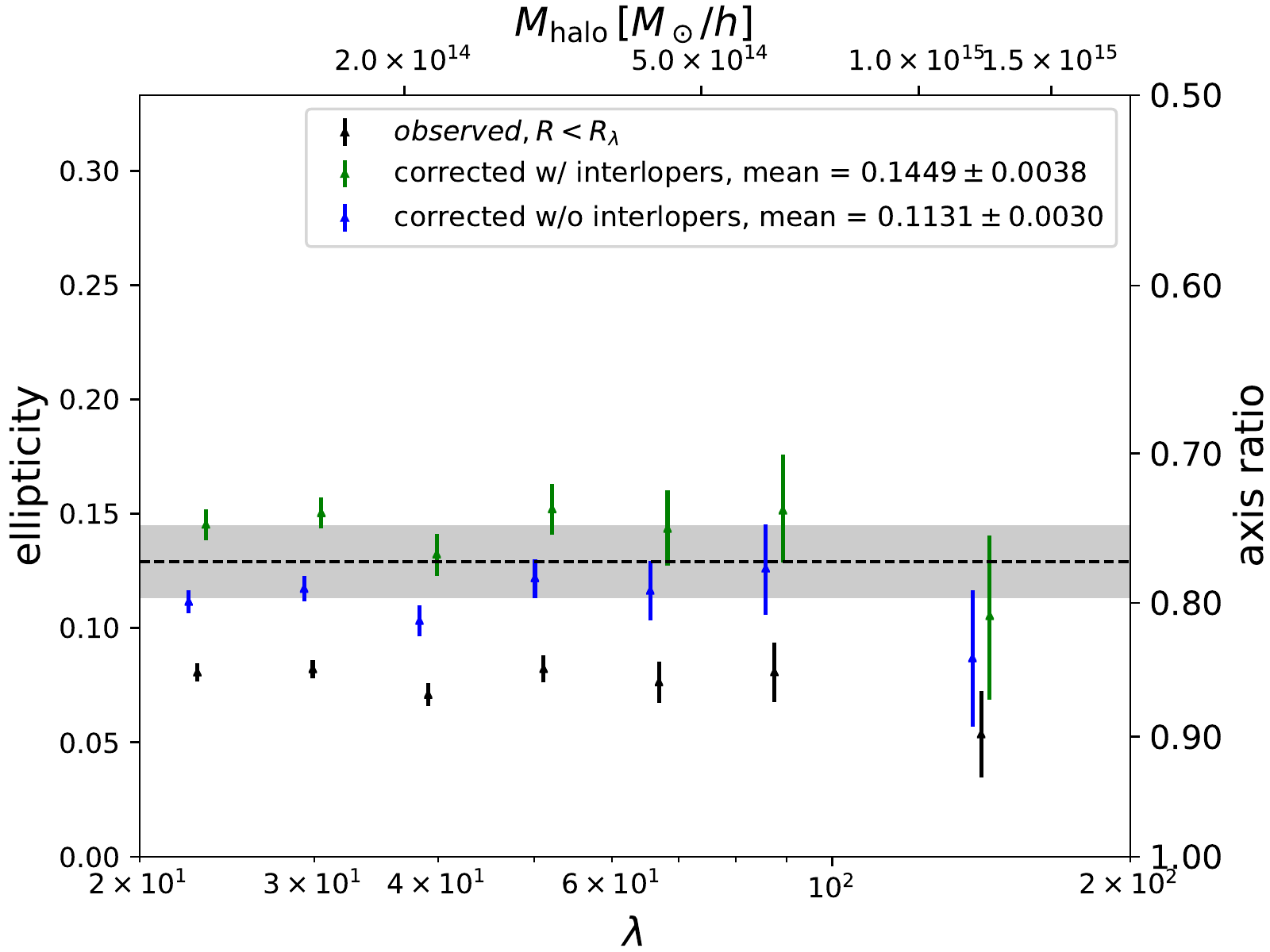}}
\caption{Ellipticity estimates of the satellite distribution as in Fig.~\ref{fig:sim-correct}, but orienting the stacked clusters along the CG major axis.
When orienting along the CG axis we are only subject to edge bias. The lower ellipticity is due to misalignment between the CG and the satellites as discussed in the text. 
}
\label{fig:sim-correct-bcg}
\end{figure}
%%%%%%%%%%%%%%%

%%%%%%%%%%%%%%%
\begin{figure}
\centering
\resizebox{85mm}{!}{\includegraphics{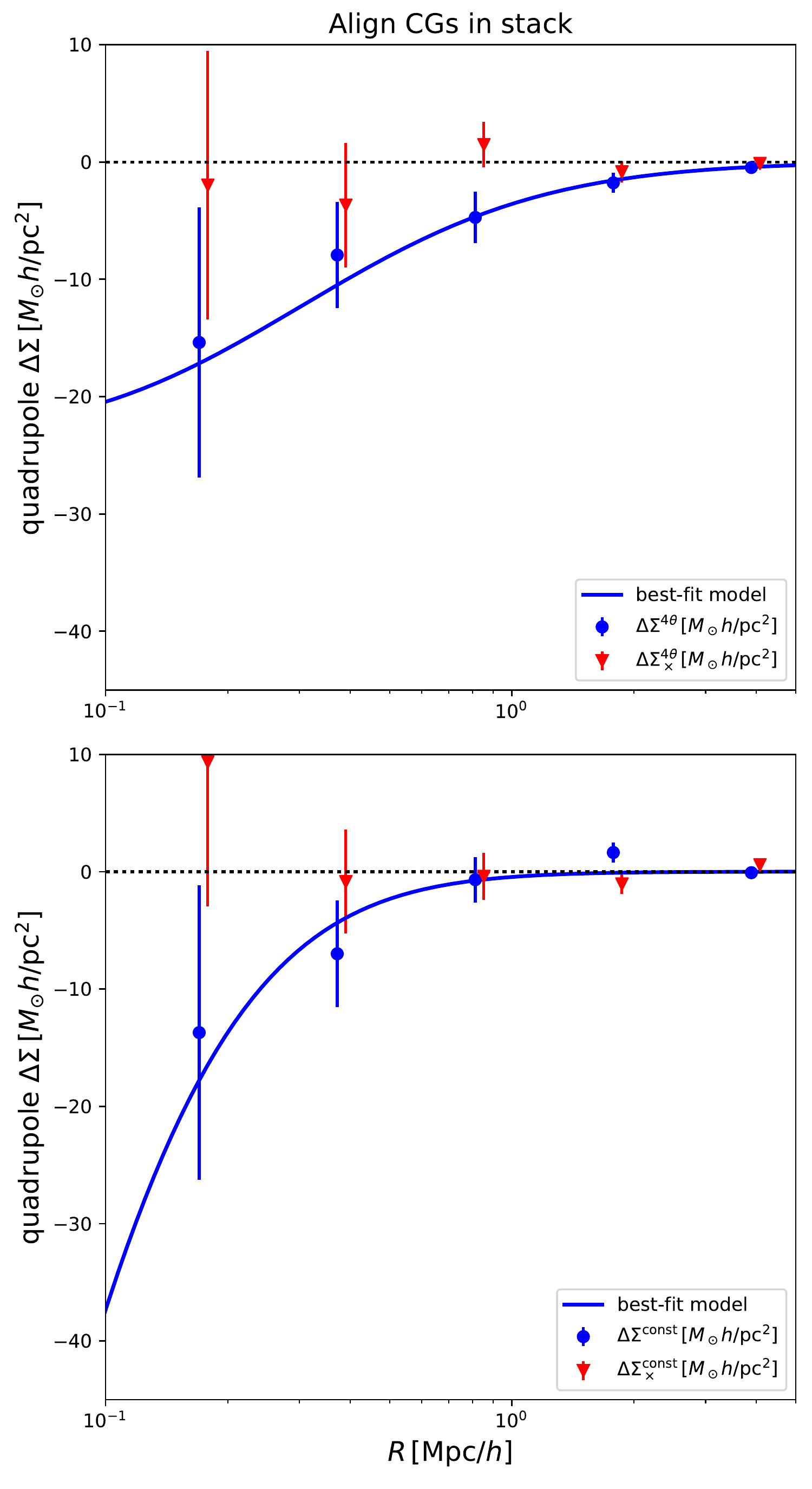}}
\caption{Same as Fig.~\ref{fig:lensing-sig-sat} but aligning the stacked clusters along the CG major axis.
The signal indicates a best-fit axis ratio of $q = 0.746 \pm 0.058$, while the cross-components are consistent with zero.
As with the satellite distribution shown in Fig.~\ref{fig:sim-correct-bcg}, the lower stacked halo ellipticity is consistent with an RMS misalignment of the CG of $\sim 30^\circ$.  
}
\label{fig:lensing-sig-bcg}
\end{figure}
%%%%%%%%%%%%%%%

%%%%%%%%%%%%%%%
\subsection{Stacking on CG major axis}
\label{sec:cg-stack}
%%%%%%%%%%%%%%%

In Fig.~\ref{fig:sim-correct-bcg} we show the ellipticity of the cluster satellite distribution when stacking along the CG major axis.
As in Fig.~\ref{fig:sim-correct} the result is shown in bins of richness both before and after correction of edge bias.
Unlike the case when stacking along the major axis of the satellite distribution itself, in this case the ellipticity of the satellite distribution is not subject to noise bias.
Noise bias only arises when the same observable is used to orient the stack and to calculate the ellipticity.

To check that the 6,681 clusters we have with good CG shape measurements are an unbiased subsample, we measure ellipticity of satellite distribution (Sec.~\ref{sec:satellite}) using only those 6,681 clusters.
We have confirmed that this selection alters the ellipticity of satellite distribution (0.271) at a sub-percent level, much smaller than our statistical error of $\sim 1\%$.

\begin{figure}
\centering
\resizebox{85mm}{!}{\includegraphics{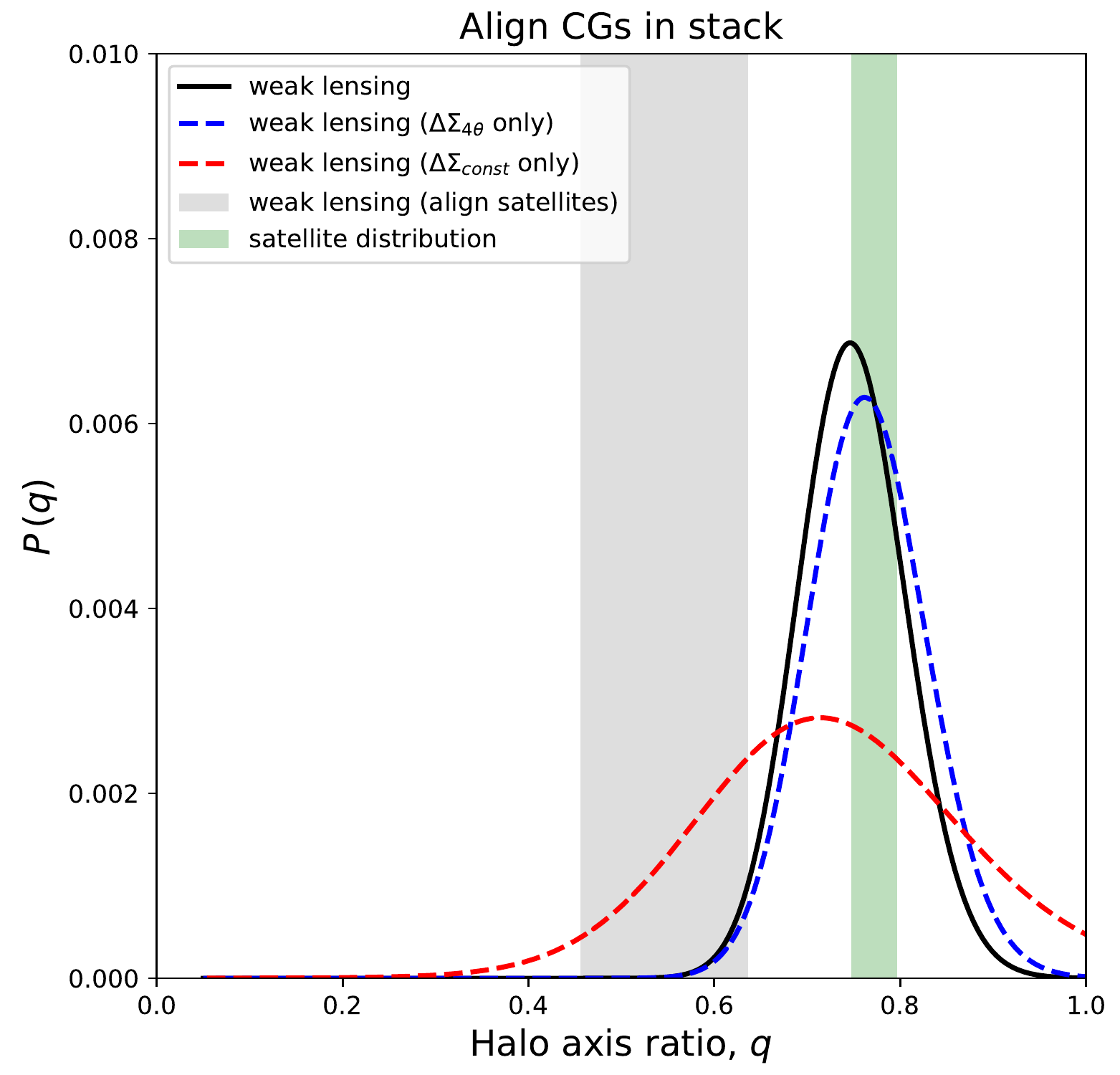}}
\caption{Same as Fig.~\ref{fig:lensing-fit-sat} except here the stacked lensing measurement is oriented along the major axis of the CG for each cluster.
The estimated axis ratios of the satellite distribution (green shaded band) and of lensing (black, blue and red lines) are consistent.
The wider, grey shaded band shows the $1\sigma$ range of the lensing measurement stacked along major axes of satellite distribution -- this is identical to the 1$\sigma$ range of the axis ratio in Figure 6. Its smaller axis ratio indicates the misalignment between the CG relative to both the satellite distribution and the lensing quadrupole.}
\label{fig:lensing-fit-bcg}
\end{figure}

In Fig.~\ref{fig:lensing-sig-bcg} we show the lensing signal when stacking on the CG major axis.
Again the signal components are consistent with coming from halo ellipticity, with a total significance of $4.4\sigma$.
This is slightly less S/N than the satellite-stacked result in Fig.~\ref{fig:lensing-sig-sat}.
There are two reasons for this: the number of clusters is smaller, increasing the noise, while we have not included the misalignment factor, decreasing the signal.
The full posterior of the lensing result is shown in Fig.~\ref{fig:lensing-fit-bcg}.
The best-fit lensing axis ratio (jointly fitting  $\dstheta$ and $\dsconst$) is $q = 0.746 \pm 0.058$, or $e = 0.145 \pm 0.038$ (black solid line).
Note that the posterior of $\dstheta$-only fitting ($0.762 \pm 0.064$, blue dashed line) and that of $\dsconst$-only fitting ($0.714 \pm 0.144$, red dashed line) are consistent to each other and with the joint-fit.
Again, in this case the satellite and lensing axis ratios agree well within $1 \sigma$ with each other.

Note that these numbers are not corrected for misalignment with the dark matter halo.
Our simple MC simulations cannot be used to correct this misalignment -- more complex hydrodynamic simulations which include the formation of the stars of the CG would be required.
However, we can estimate the misalignment of the CG by calculating how much dilution would be required to match the CG-stacked ellipticity to the satellite-stacked ellipticity.
This corresponds to a root-mean-squre (RMS) misalignment angle of $30^\circ \pm 10^\circ$ between the central galaxy and dark matter halo.
This is a slightly larger misalignment than that of satellite galaxies, which have an RMS misalignment of $18^\circ \pm 2.5^\circ$.
Moreover the CG case is a `true' misalignment as the measurement error in its orientation is small, while for satellite galaxies the misalignment is due to Poisson sampling.

%%%%%%%%%%%%%%%
\section{Comparison with other work}
\label{sec:compare}
%%%%%%%%%%%%%%%

%%%%%%%%%%%%%%%
%%%%%%%%%%%%%%%

%%%%%%%%%%%%%%%
\subsection{Simulations}
%%%%%%%%%%%%%%%

\citet{despali17} studies halo shapes in the Le SBARBINE N-body simulations \citep{despali16}.
They use an Ellipsoidal Overdensity halo finder which is similar to Spherical Overdensity finders except without imposing a spherical shape.
Their Figure 11 shows 2D, projected halo axis ratios.
Our SDSS sample covers a range of halo masses between $10^{14}$ and $10^{15} \msun/h$, at $z \sim 0.3$.
According to their Figure 11, halos with these masses have axis ratios of 0.65 and 0.55, respectively.
This gives an average of approximately 0.6 for our sample.
This value is within $1\sigma$ of the lensing result shown in Table 1, $0.558 \pm 0.086$.
It is also within $1\sigma$ of the satellite result, $0.573 \pm 0.039$.

The results of \citet{despali17} are consistent with previous N-body simulations, for example, the 2D projected ellipticity plots in \citet{hbb06}.
This work found axis ratios of 0.645 (0.621) for halos above $10^{14} M_\odot / h$ at $z=0.25$ for cosmologies with $\Omega_m = 0.3$ and $\sigma_8 = 0.9 \, (0.7)$. 
This axis ratio also agrees with our result at the $\sim 1\sigma$ level.
(Note that \citet{hbb06} used a different ellipticity definition, $q = (1-e)^2$. We have obtained the axis ratio numbers via their Figure 1 which shows ellipticities $\sim 0.197 \, (.212)$ for these two cases.)

\citet{velliscig15} show results for halo axis ratios in hydrodynamical simulations.
Results were obtained over a wide range in halo mass, $10^{10} - 10^{15} \msun/h$, by using multiple simulations with different resolutions and sizes.
(Small, high-resolution simulations are used to study small halos while large, lower-resolution simulations study rarer, more massive halos.)
Reading off their Fig. 3, they find average 3D (not projected) axis ratio $\sim 0.55$ for clusters with mass of $\sim 10^{14} \msun/h$.
We can approximately convert this 3D axis ratio into a 2D axis ratio $\sim$ 0.7 using Fig. 9 of \citet{despali17}.
The inferred 2D axis ratio in \citet{velliscig15} is larger at about 2$\sigma$ than our lensing axis ratio $= 0.558 \pm 0.086$ -- further detailed analysis is needed to identify possible sources of this discrepancy.

%%%%%%%%%%%%%%%
\subsection{Observations}
%%%%%%%%%%%%%%%

Observational studies of cluster shapes include both the satellite distribution and lensing measurements. A detailed 
and recent study is that of \citet{van16} who measure the ellipticities of 2600 groups ($\lambda > 5$) over 180 square degrees from the GAMA group catalog of \citet{robotham11}.
(Note that the \citet{van16} groups are smaller than our clusters: only $\sim$ 100-200 of the objects in their sample have richness $\lambda > 20$.)
They use the KiDS weak lensing catalogs of \citet{hildebrandt17} to measure the lensing quadrupole.
Stacking on the CG major axis, they find a $3.2 \sigma$ detection of lensing ellipticity assuming a diagonal covariance matrix. (Using an alternative estimator that  cancels the spurious alignment between lenses and sources, the signal strength is reduced to $2.5 \sigma$; see Sec.~4.1 of that work.)
Stacking the lensing measurement on the BCG major axis, their best-fit value is $e = 0.38 \pm 0.12$, corresponding to an axis ratio $= 0.45^{+0.14}_{-0.12}$ over 40 kpc $< R < 250$ kpc.
While this result is consistent with our result, it should not be directly compared with ours because \citet{van16} 
assumes a perfect alignment between the visible light (CG or satellites) and dark matter major axes.

Stacking along the major axis of the satellite galaxy distribution, \citet{van16}  find $e = 0.49 \pm 0.13$.
In their Eq. (37) they estimate the dilution due to misalignment as $\sim 20-25\%$ when stacking on the satellite major axis.
Revising their result to include this misalignment results in $e = 0.65 \pm 0.17$ or axis ratio $=0.21^{+0.14}_{-0.11}$.
This is a $\sim$2$\sigma$ tension with our result, $0.558 \pm 0.086 {\rm (stat)} \pm 0.026 {\rm (sys)}$. It is also a nearly $3\sigma$ tension with the \citet{despali17} N-body results, which obtain a 2D axis ratio of $\sim 0.65$ for group-size halos with mass $M\sim 10^{14} \msun/h$.
Halo mass dependence does not resolve this tension as the more massive halos in the N-body results should be more elliptical, not less.
Note that their assumption of a diagonal covariance also cannot explain this discrepancy since we do not see any strong off-diagonal component of our covariance.

One possible source for the higher ellipticity is due to the projected galaxies that are not in fact cluster members (for example see \citealt{farahi16}) and impact the lower richness groups more. 
These ``accidentally'' projected galaxies are more likely to lie on the outskirts of the lens halos and could bias the major axes  toward their position.
Then stacking the groups along the calculated major axes would cause a spurious quadrupole lensing detection at large radii. 
 Table 1 in \citet{van16} indeed shows that stacking along the major axis of satellite distribution leads to a $\sim 0$ ellipticity at small radii (40 kpc $< R < 250$ kpc) but a large ellipticity of $e = 0.49 \pm 0.13$ for 250 kpc $< R < 750$ kpc. 

The earlier study of \citet{evans09} extends to slightly larger groups and clusters in SDSS.
Their sample goes down to $\lambda = 10$ with a mean richness $\sim 10^{14} M_\odot / h$.
They find a best-fit axis ratio of $0.48^{+0.14}_{-0.09}$, assuming an NFW model.
For the galaxy distribution, they find an axis ratio of 0.6.
These numbers are within $1\sigma$ of \citet{despali17} and our results, although again cannot be compared directly as they   do not include a  correction for the Poisson sampling of the satellites.

\citet{huang16} found an RMS misalignment of $35.07^\circ \pm 0.28^\circ$ between the major axis of the satellite distribution and of the CG light.
Assuming the misalignment between DM and CG major axes is independent of the misalignment  between DM and the satellite distribution, we find an RMS misalignment between the major axis from the satellite distribution and the CG light to be $(30^{2} + 18^{2})^{0.5} = 35 \pm 10$ degrees. Our result is consistent with the result from \citet{huang16}.

%%%%%%%%%%%%%%%
\section{Discussion}
%%%%%%%%%%%%%%%

We have measured the ellipticity of SDSS clusters stacked along the major axes of their satellite distribution and those of the CGs. 
We have used two different methods (ellipticity of the satellite distribution and ellipticity from quadrupole lensing), and found that the results from the two methods are consistent with each other.

To measure the ellipticity of the satellite distribution, we began by modeling the clusters with MC simulations. 
The model includes satellites which follow an elliptical NFW profile and interlopers which are uniformly distributed.
This model allows us to correct for several biases in the raw ellipticity measurements.
First, Poisson noise, due to the finite number of member galaxies in a cluster, causes the raw measured ellipticity to be larger than its true value.
Second, the \redmap algorithm's hard edge (at $R=R_{\rm \lambda}$) can decrease measured ellipticity by excluding galaxies along the cluster's major axis with $R>R_{\rm \lambda}$.
We believe we have modeled and corrected these biases sufficiently accurately so that they do not contribute to our systematic uncertainty, but interlopers are more difficult to model.
To estimate the error due to interlopers, we considered two extreme cases: one with a large interloper fraction (typically 40\%), and one with no interlopers at all.
The difference between these cases is about a $\sim 10\%$ uncertainty in ellipticity, which is a conservative estimate of the effect of interlopers.
The raw and corrected ellipticity measurements for both cases are shown in Fig.~\ref{fig:sim-correct}.
Our final result for cluster ellipticity of the satellite distribution is $0.271 \pm 0.002 ({\rm stat}) \pm 0.031 ({\rm sys})$, which corresponds to an axis ratio of $0.573 \pm 0.002 ({\rm stat}) \pm 0.039 ({\rm sys})$.

In the context of the satellite distribution we have tested a number of other effects.
These are generally less important than the biases mentioned above and are summarized in Appendix~\ref{sec:tests}.
They include ellipticity of the $\pmem$ filter (\ref{sec:ellip-pmem}),
subhalo clustering due to infall as a group (\ref{sec:group-infall}),
ellipticity variation due to intrinsic differences between halos or projection effects (\ref{sec:ellip-dist}), and
richness errors in the \redmap algorithm (\ref{sec:richness-error}).
In addition, there might be contamination from projected foreground structures which get included in the richness estimation.
In this case, it is possible that the fraction of interlopers is larger than assumed here. We try to minimize this effect by selecting member galaxies with $\pmem>0.5$ (we checked higher values as well, as discussed in Appendix B).
\citet{farahi16} estimates the ``total'' non-member fraction, including both the non-member fraction that is imprinted in $\pmem$ values by $\sum {(1 - \pmem)}$ and the non-member fraction that is included in the richness estimation, in SDSS \redmap clusters in their Table 2 to be $\sim 0.4$. This is smaller than our average non-member fraction $1-<\pmem>=0.473$, therefore, adding more interlopers to account for this projection effect appears unnecessary.

So far, we have assumed that the lensing signal would be dominated by dark matter. However,  rich clusters can have up to $\sim 15\%$ of the mass in hot gas, which can have a non-negligible effect on the lensing signal. If the gas is not well aligned with the shape of the dark matter halo, it would alter the ellipticity. Accounting for this effect is out of the scope of this work and we leave it to further studies using the results of simulations that include gas physics. We do compare with one study of AGN feedback below.

The second major aim of this paper is a measurement of the lensing quadrupole of cluster halos. We presented new optimal estimators which allowed a $5\sigma$ measurement of the cluster lensing quadrupole.
Aligning the stacked lensing measurement with the satellite major axis, we found a mean axis ratio of $q = 0.558 \pm 0.086$(stat) $\pm 0.026$(sys), in good agreement with the axis ratio from the satellite distribution itself.
This axis ratio, which has already been corrected for Poisson sampling of satellites, 
represents our constraints on the true {\it dark matter} halo axis ratios of \redmap clusters.
Thus it can be directly compared to cluster-sized halos in N-body simulations -- the agreement to the N-body results is within $1\sigma$ (see Sec.~\ref{sec:compare}). 

We repeated all satellite and weak lensing axis ratio measurements using the central galaxy (CG) light's major axis to orient the stack.
Again, agreement between the satellite ($q = 0.772 \pm 0.025$) and weak lensing ($q=0.746 \pm 0.058$) axis ratio is within $1 \sigma$.
Here the lensing detection was slightly weaker at $4.4\sigma$ due to the smaller number of clusters available (about 1/3 of CGs had poor shape measurements) and greater observable-halo misalignment. We infer that the CG is misaligned with respect to the cluster satellite distribution by an RMS angle of $35^\circ \pm 10^\circ$, consistent with Huang et al. (2016).
These two observable methods for orienting the stack are subject to different systematics.
For example, the CG and nearby background sources can exhibit spurious alignments \citep{mhb06, clampitt16} due to PSF- and camera-based effects.
The consistency between the ellipticity from satellite distribution and lensing using both observables to orient the clusters supports the robustness of our measurements.

One application of halo ellipticity measurements is to constrain models of baryonic physics.
Our mean axis ratio of $q = 0.558$ is in some tension with the no-AGN simulations of \citet{suto16}.
Based on their Figure 9 the integrated probability (DM curve, left panel) from 0 to $q = 0.558$ is $\sim$ a few percent.
In contrast, the simulations with AGN (right panel) show consistency within $\sim 1\sigma$ with our results.
However, it is difficult to say with certainty due to the small number of clusters (40) in the \citet{suto16} simulations as well as their somewhat smaller mass ($> 5\times 10^{13} M_\odot/h$) and the overcooling problem in their no-AGN simulations. Based on \citet{despali17} the shift in axis ratio due to the relatively lower mass cutoff is fairly small ($\sim 0.05$).
More precise constraints on the models with AGN or other feedback will need to wait for larger simulations. Such simulations are underway by the Eagle (\citet{barnes17}) and Illustris (\citet{nelson15}) collaborations. 

In  future work it will be interesting to test for variation in halo ellipticity with radius. The effects of AGN feedback are expected to dominate the inner parts of the cluster, as are those of self-interacting dark matter which makes the inner density contours rounder (see \citet{brinckmann17} for a recent study from simulations). It will be a challenge to simultaneously test for these and other effects that go beyond the baseline CDM model, but the joint analysis of the monopole and quadrupole as a function of scale and other parameters should provide useful test. 
With current data the radial dependence is difficult to test because of the way edge bias interacts with $\pmem$ (see Appendix~\ref{sec:app-pmem}).
Other work, for example \citet{zu16}, using \redmap clusters have also noted difficulties from selection effects produced by cuts on $\pmem$.
Finally, studies of the alignment of the dark matter halo with large scale structure (LSS) can extend current work comparing light and LSS alignments.
For example, \citet{zhang13} has shown, using SDSS DR7 data, that galaxies within filaments are aligned with the filament direction.
The alignment is strongest for red, central galaxies of groups or clusters.
It will be interesting to see whether the alignment between halos and filaments are even stronger.
With ongoing surveys such as the Dark Energy Survey (\citet{des05}) we expect to have at least a factor of two improvement in the statistical signal-to-noise (which dominates for lensing). Future surveys from LSST (\citet{abell09,ivezic08}), Euclid (\citet{laureijs10}) and WFIRST (\citet{spergel13}) will provide huge gains in statistical accuracy and the opportunity to study trends with radius, richness and redshift. 

\section*{Acknowledgments}
We would like to thank Eric Baxter, Joe DeRose, Ian Dell'Anotonio, Daniel Gruen, Hung-Jin Huang, Benjamin Joachimi, Andrey Kravtsov, Rachel Mandelbaum, Surhud More, Yuedong Fang, Mike Jarvis, Masahiro Takada, Joop Schaye, Frank van den Bosch, Mark Vogelsberger, Anja von der Linden and Risa Wechsler for helpful discussions.
We are very grateful to Erin Sheldon for allowing us to use his SDSS shear catalogs and source photometric redshift distributions. JC, BJ and GB are partially supported by the US Department of Energy grant DE-SC0007901 and funds from the University of Pennsylvania.

%%%%%%%%%%%%%%%%%%%%%%%%%%%%%%%%%%%%%%%%%%%%%%%%%%



%\bibliographystyle{mnras}


%%%%%%%%%%%%%%%%%%%%%%%%%%%%%%%%%%%%%%%%%%%%%%%%%%

%%%%%%%%%%%%%%%%% APPENDICES %%%%%%%%%%%%%%%%%%%%%
\appendix

%%%%%%%%%%%%%%%%
\section{Effect of Different $\pmem$ Cuts}
\label{sec:app-pmem}

\begin{figure*}
\centering
\resizebox{55mm}{!}{\includegraphics{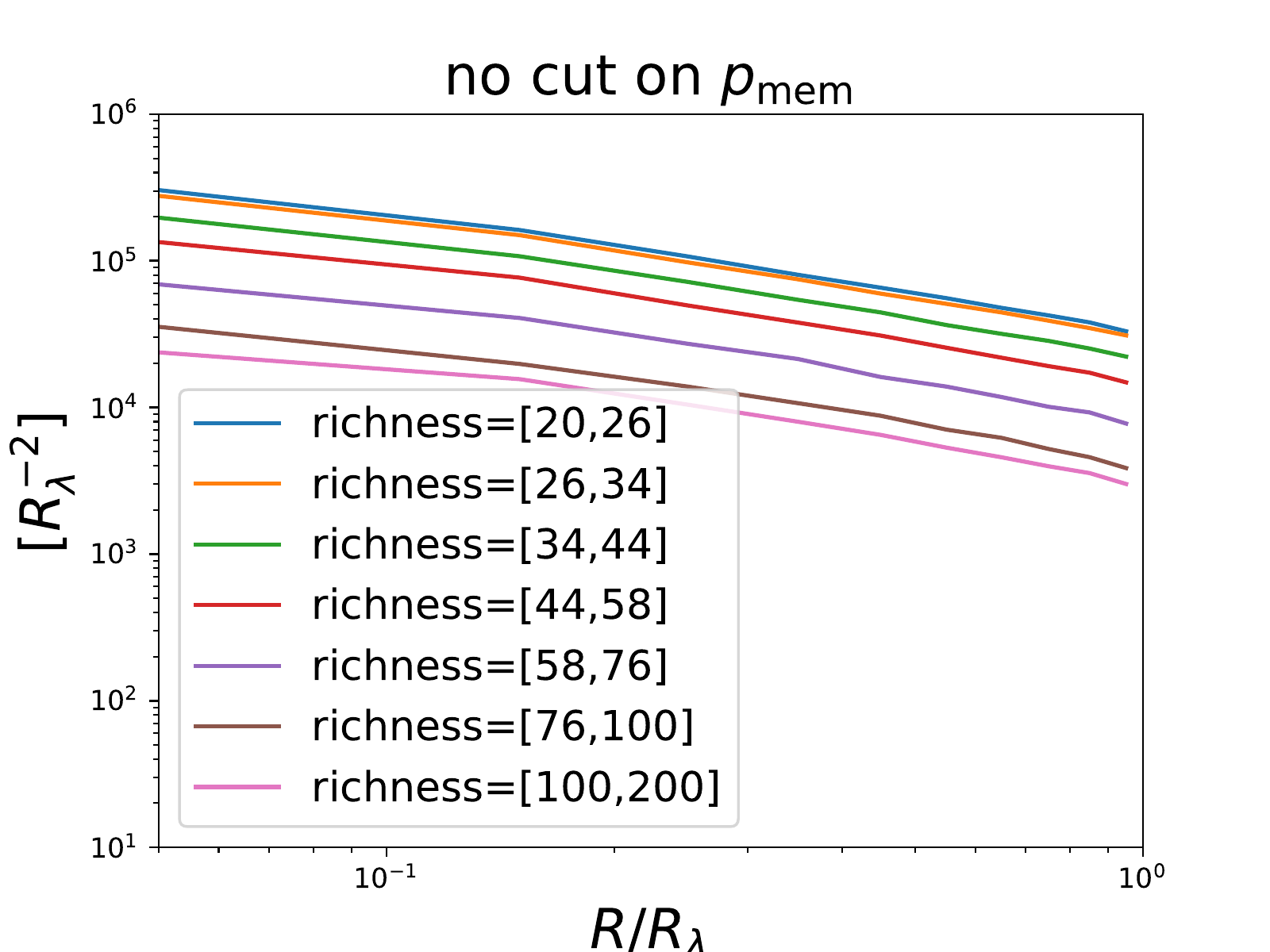}}
\resizebox{55mm}{!}{\includegraphics{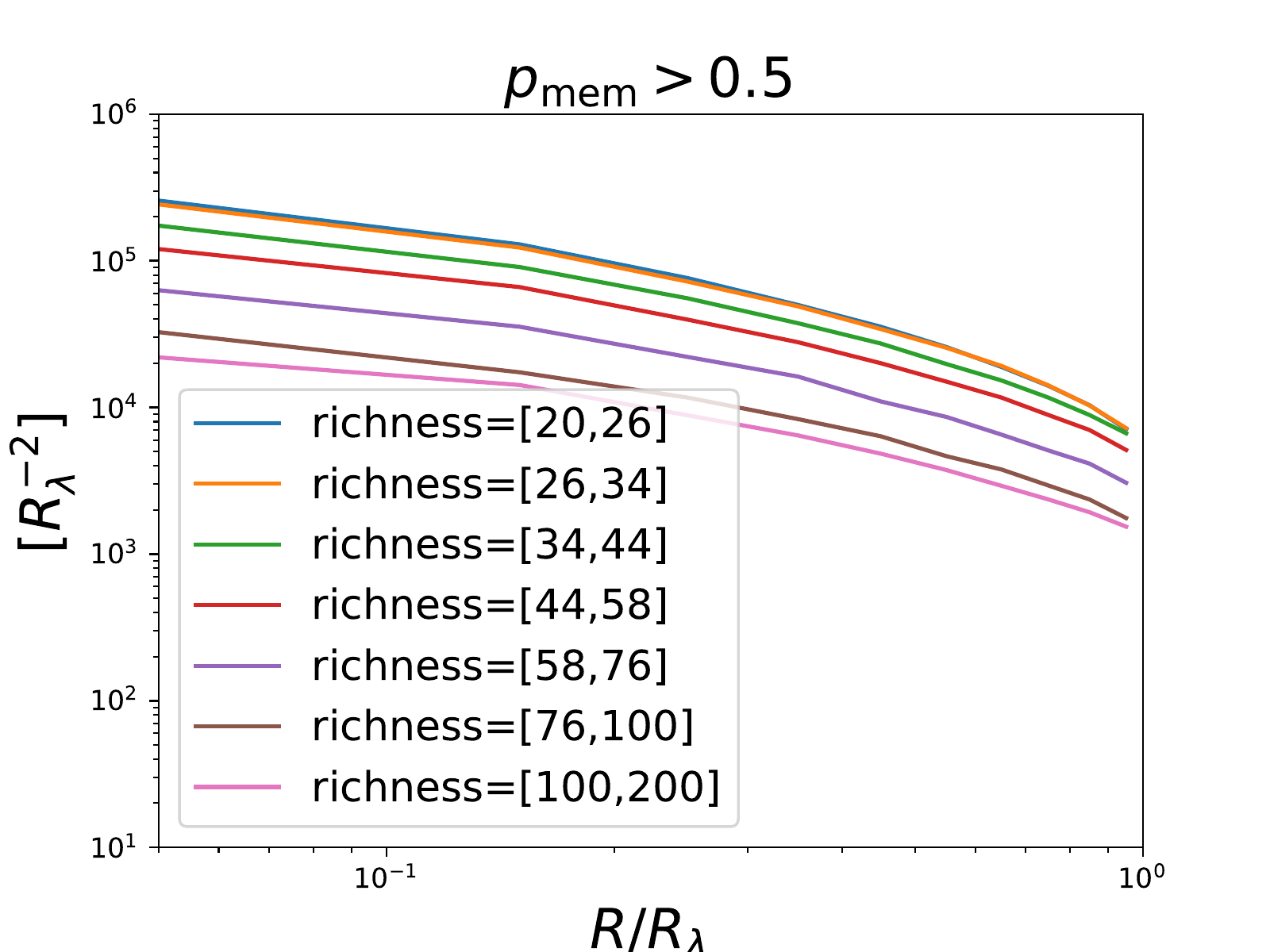}}
\resizebox{55mm}{!}{\includegraphics{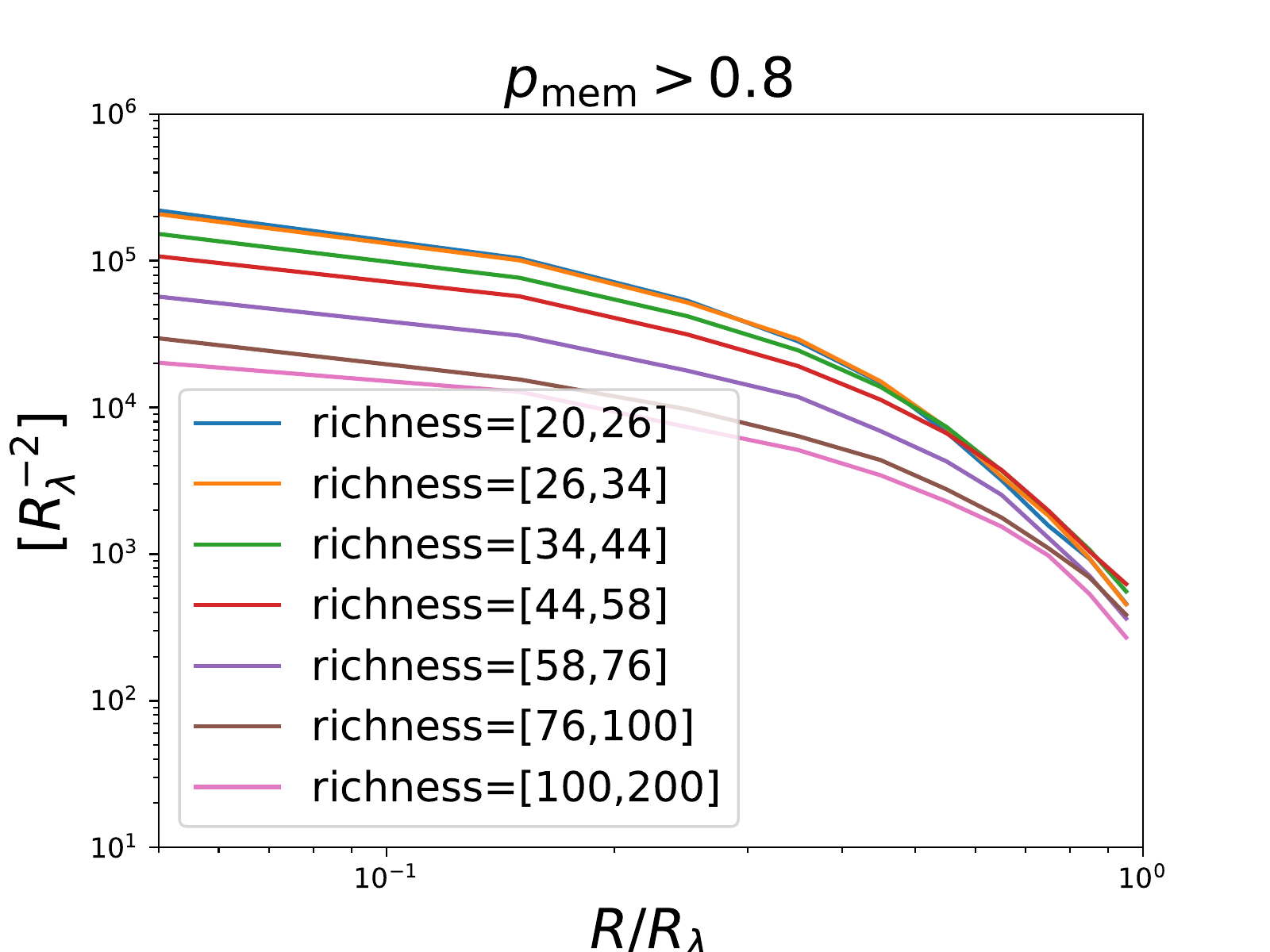}}
\caption{Stacked number density of \redmap clusters in each richness bin.
({\it left panel}): Without any cut on $\pmem$.
({\it middle panel}): With $\pmem$>0.5.
({\it right panel}): With $\pmem$>0.8.
The number of member galaxies in the outermost region decreases significantly as we apply a stricter $\pmem$ cut.}
\label{fig:num-den}
\end{figure*}

For the edge bias correction (Sec. ~\ref{sec:edge-bias}) to be accurate, we need sufficiently high galaxy number density around the edges, $R \sim R_{\rm \lambda}$, of \redmap clusters.
With too aggressive a $\pmem$ cut, the density profile contracts, falling off faster than NFW.
It then fails to match our MC simulation with elliptical NFW out to $R = R_{\rm \lambda}$.
In Fig.~\ref{fig:num-den}, we compare the stacked number density profiles with different $\pmem$ cuts.
For example, implementing $\pmem>0.8$ leaves 10-50 times fewer member galaxies on the outskirts compared to the case with no $\pmem$ cut at all.
Considering the total number of member galaxies per cluster (order of $10-100$) and the decreasing tendency of $\pmem$ with projected distance from the cluster centre, this suggests that there will be almost zero member galaxy left on the outskirts. 
On the other hand, a $\pmem>0.5$ cut leaves about 25-60\% of member galaxies in the outermost region.
This better approximates a sharp boundary at $R=R_{\rm \lambda}$ as in our MC simulations.
Therefore, we decide to use the $\pmem>0.5$ cut as our fiducial case, since further relaxing the cut to a lower value of $\pmem$ causes higher systematic error in ellipticity due to increased interlopers.
For instance, if we do not apply any $\pmem$ cut at all, the uncertainty on ellipticity increases by $\sim 150\%$, compared to the $\pmem$>0.5 case, while the value of ellipticity stays at $\sim0.27$.

Furthermore, the corrected value of ellipticity is not very sensitive to the choice of $\pmem$ cut.
We have checked that using different criteria for the $\pmem$ cut makes less than a 5\% difference on the measured ellipticity value.

These uncertainties due to interlopers and $\pmem$ cuts are the most significant, so we have emphasized them here in the main text.
In Appendix~\ref{sec:tests} we describe a few other potential sources of error that turned out to be negligible.

%%%%%%%%%%%%%%%%%%%%%
%%%%%%%%%%%%%%%%%%%%

%%%%%%%%%%%%%%%
\section{Miscellaneous tests}
\label{sec:tests}
%%%%%%%%%%%%%%%

%%%%%%%%%%%%%%%
\subsection{Elliptical $\pmem$ distribution}
\label{sec:ellip-pmem}

The value of $\pmem$ strongly depends on the projected distance from the cluster centre. However, it could also have an angular dependence.
In this case, the soft edge that a $\pmem$ cut imposes becomes elliptical, engendering totally a different edge bias.
Because the satellite distribution is typically misaligned with the underlying dark matter distribution, we cannot figure out how elliptical this additinoal soft edge (by a $\pmem$ cut) of the stacked clusters would be, unless accurate an N-body simulation is provided.
We calculate the monopole signal of this soft edge that our $\pmem$ cut imposes by dividing the average of $\pmem>0.5$ by that without any $\pmem$ cut, as a function of projected distance from the cluster centre with an appropriate normalization.
This monopole can be used to construct two extreme cases to estimate the possible systematic due to this uncertainty in the ellipticity of $\pmem$ distribution : 1) perfectly circular edge and 2) elliptical edge with a specific ellipticity we want to apply.
We can apply these two extreme cases of soft edges to our MC simulation and measure the true ellipticity of each case.
For example, we measure the true ellipticity of 0.40 assuming a circular edge, whereas we measure the true ellipticity of 0.18 assuming an elliptical edge with ellipticity of 0.3.

\subsection{Satellite infall in groups or filaments}
\label{sec:group-infall}

We know that these \redmap members are clustered with each other \citep{fang16}.
Our simulations don't include this clustering of red galaxies, so they will slightly overestimate ellipticity by undercorrecting for noise bias.
We estimate the size of this effect based on the measurements of \citet{fang16}.
Using the definition of the angular correlation function $w$, for an annulus at a projected distance $R$ with area $dA$, the number of satellites around a given satellite is $\de N (R) = \de A \times \bar{n}_{nfw}(R)\times [1+w(R)]$, where $\bar{n}_{nfw}(R)$ denotes a smooth, unclumped NFW profile.
Therefore, the excess due to correlations at a projected distance $R$ is just $w(R)$.
If we integrate that over the cluster, centered on a typical satellite at $\sim 0.3 \mpch$ from the cluster centre, then we get the fractional excess in counts (and hence overestimate of noise bias) due to satellite clustering.
 
We can estimate this quantity using Figure 1 of \citet{fang16}.
The ratio of satellite correlations (red points) to the correlations associated with the host cluster (black points) ranges from 0.2 to 0 over $R = 0.1 - 1 \mpch$.
Since the average $\bar{w}(R) \sim \int R \, \de R \, w(R)$ is about 0.1, we would need to correct noise bias by a factor of 1.1.
In other words, satellite correlations mean that there are 10 percent fewer independent points.
Note that \citet{fang16} showed cross-correlations of cluster members with \redmagic galaxies, while we are interested in the auto-correlations of cluster members.
However, the results are the same: we have checked that the ratio of auto-correlations of cluster members to cluster-member correlations is the same as the analogous ratio shown in Figure 1 of \citet{fang16}.

Next, we compare this estimate of 10\% fewer independent points to the noise bias correction in Fig.~\ref{fig:app-fake}.
From Fig.~\ref{fig:app-fake} we estimate the effect of 10\% fewer independent points is negligible.
Noise bias varies most rapidly for smaller $\lambda$, so that's where this effect will have the biggest impact. 
Focusing then on the difference between the $\lambda=26-34$ bin and the 20-26 bin we see the difference between these bins is about 5\% of the true ellipticity (left panel, for the ellipticity values $\sim 0.25$).
But this is an overestimate, since 10\% of lambda $\sim 30$ is just a shift of 3.
So we have, at worst, $\sim 3\%$ bias in ellipticity.
The uncertainty in background galaxies gives us an $\sim 10\%$ systematic uncertainty (see Sec.~\ref{sec:fake-member}).
So the error of $\lesssim$ 3\% from satellite clustering is much smaller and we neglect it.

%%%%%%%%%%%%%%%
\subsection{Broad ellipticity distribution}
\label{sec:ellip-dist}

In Sec.~\ref{sec:satellite} we use MC simulations to correct for biases in ellipticity of satellite distribution.
In doing so, we swept all variations of ellipticity away into an average correction factor.
Such variations are expected from (i) intrinsic differences between halos (some are more elliptical than others) and (ii) differences in their orientiation with respect to the line-of-sight.
Thus, our noise and edge bias corrections could be misestimated if the distribution of ellipticity is very broad or highly skewed.
While the corrections in Figures~\ref{fig:app-nofake} and \ref{fig:app-fake} are fairly smooth functions of input ellipticity, we perform an additional test to verify the accuracy of our simulations.

\begin{figure}
\centering
\resizebox{90mm}{!}{\includegraphics{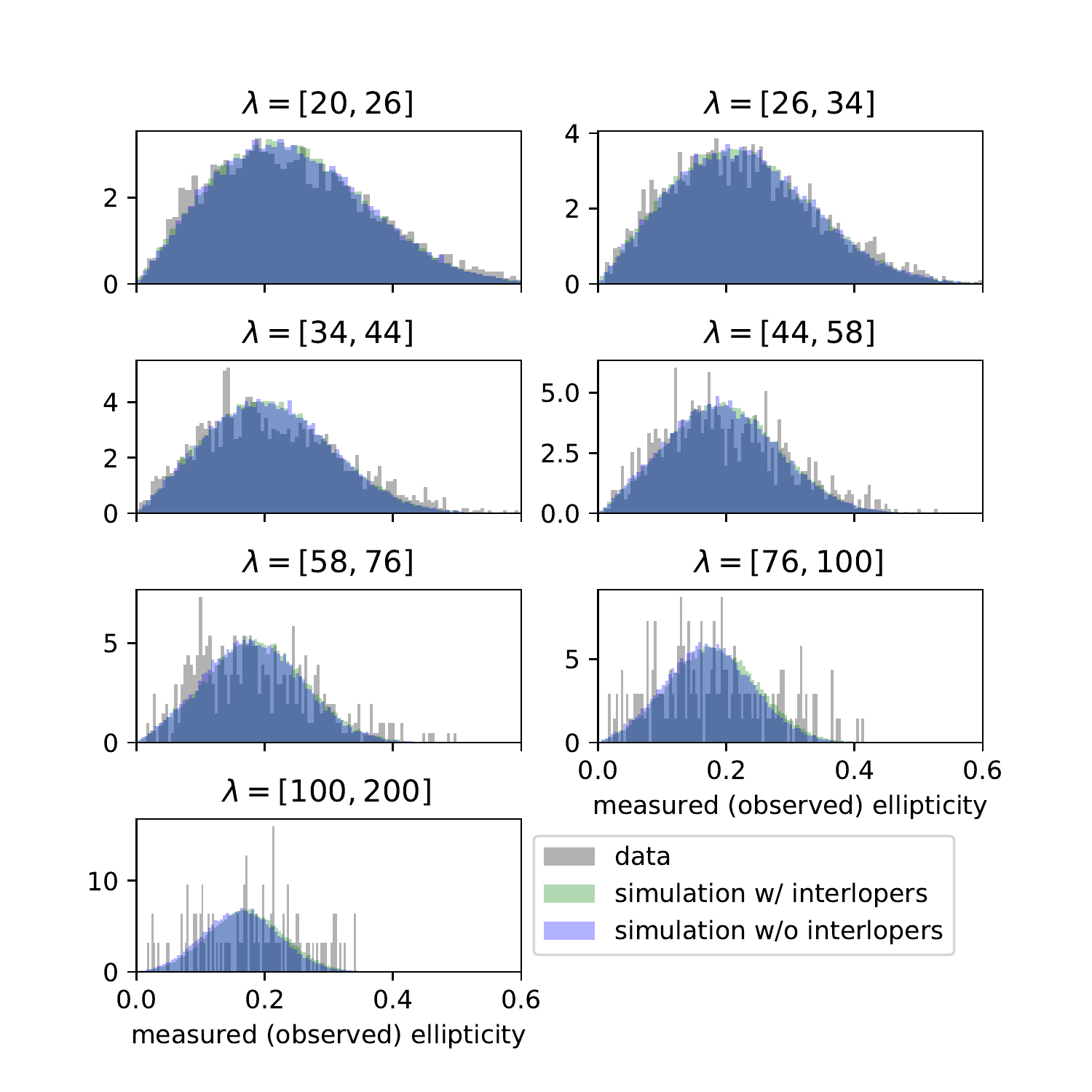}}
\caption{Ellipticity distribution of each richness bin from data and simulations. ({\it black (A)}): raw ellipticity distribution of the data before bias corrections (black points in Fig.~\ref{fig:sim-correct}, see Sec.~\ref{sec:method}). ({\it green (B)}): ellipticity distribution of the simulation with interloper galaxies assuming ${\rm e_{true}} = 0.303$, the retrieved true ellipticity when interlopers are included (green points in Fig.~\ref{fig:sim-correct}, see Sec.~\ref{sec:fake-member}). ({\it blue (C)}): same as B, but excluding interlopers in the simulation and taking ${\rm e_{true}} = 0.240$, true ellipticity retrieved when interlopers are not included (blue points in Fig.~\ref{fig:sim-correct}, see Sec.~\ref{sec:result-bias}). 
The three distributions agree well with each other, therefore, using the average value of the ellipticity would not bring about any significant bias.}
\label{fig:ell-dist}
\end{figure}

If the ellipticity distributions of simulation and the data are similar to each other, using the average value will not result in any bias.
Thus, we compare distribution of raw ellipticity of data to that of our MC simulation (Sec.~\ref{sec:satellite}) in each richness bin.
We take ellipticity distributions from the same MC simulation described in Sec.~\ref{sec:satellite}, before average ellipticity values are calculated. 
The results are shown in Fig.~\ref{fig:ell-dist}.
We compare three distributions, (a) raw ellipticity measurement of data (black distribution, the black points in Fig.~\ref{fig:sim-correct} are the average values of this distribution), (b) measured ellipticity in our MC simulation with interloper galaxies considered (green distribution) and (c) measured ellipticity in the MC simulation without including any interloper (blue distribution). Note that for the comparison to be accurate, we must assume the true ellipticity retrieved in Sec.~\ref{sec:satellite}, 0.303 and 0.240 respectively for the case (b) and (c) (Fig.~\ref{fig:sim-correct}).

For the first four richness bins with $\lambda \lesssim 60$ the distributions match very well.
The ellipticity distributions of some richer cluster bins are slightly broader than or skewed relative to the data.
But these distributions are also noisier, and statistical errors on the ellipticity itself are higher.
We conclude that any additional systematic error from averaging over the broad ellipticity distribution is well below the uncertainty due to interlopers, our most important systematic.

%%%%%%%%%%%%%%%
%%%%%%%%%%%%%%%

%%%%%%%%%%%%%%%
\subsection{Impact of Richness errors on biases corrections}
\label{sec:richness-error}

Edge bias (Sec.~\ref{sec:edge-bias}) could be inaccurate due to errors in cluster richness ($\lambda$) of \redmap.
If $\lambda$ is underestimated, we consequently underestimate the size and the mass of the underlying dark matter halo.
This would cause an undercorrection of the edge bias if the edge bias correction is a strong function of $\lambda$.
However, fortunately, the edge bias is insensitive to $\lambda$. According to the middle panel of Fig.~\ref{fig:app-nofake} and \ref{fig:app-fake}, one can readily conclude that the edge bias is very weak function of $\lambda$. 
Even in an extreme case assuming that $\lambda \sim 20-26$ of a cluster is mis-estimated to $\lambda \sim 100-200$, we would only have a few percent shift on the edge bias. It is well under our systematic uncertainty of $\sim 12\%$ from the uncertainty in the number of interlopers in clusters.
Thus we conclude that richness error does not alter our conclusion.

On the other hand, noise bias is a function of number of {\it observed} member galaxies.
If unobserved galaxies were included in the calculation, we would obtain a smaller value of the raw ellipticity measurement due to the increased number of galaxies (left panels of Fig.~\ref{fig:app-nofake} and \ref{fig:app-fake}).
However, larger number of member galaxies implies smaller noise bias correction.
As a result, the same true ellipticity will be retrieved no matter how many unobserved member galaxies there are.
Since our aim is to obtain one representative value of ellipticity over all the richness range, the error on richness does not affect our result.

%%%%%%%%%%%%%%%%%%%
\section{Misalignment}
\label{sec:misalign}

\begin{figure}
\centering
\resizebox{85mm}{!}{\includegraphics{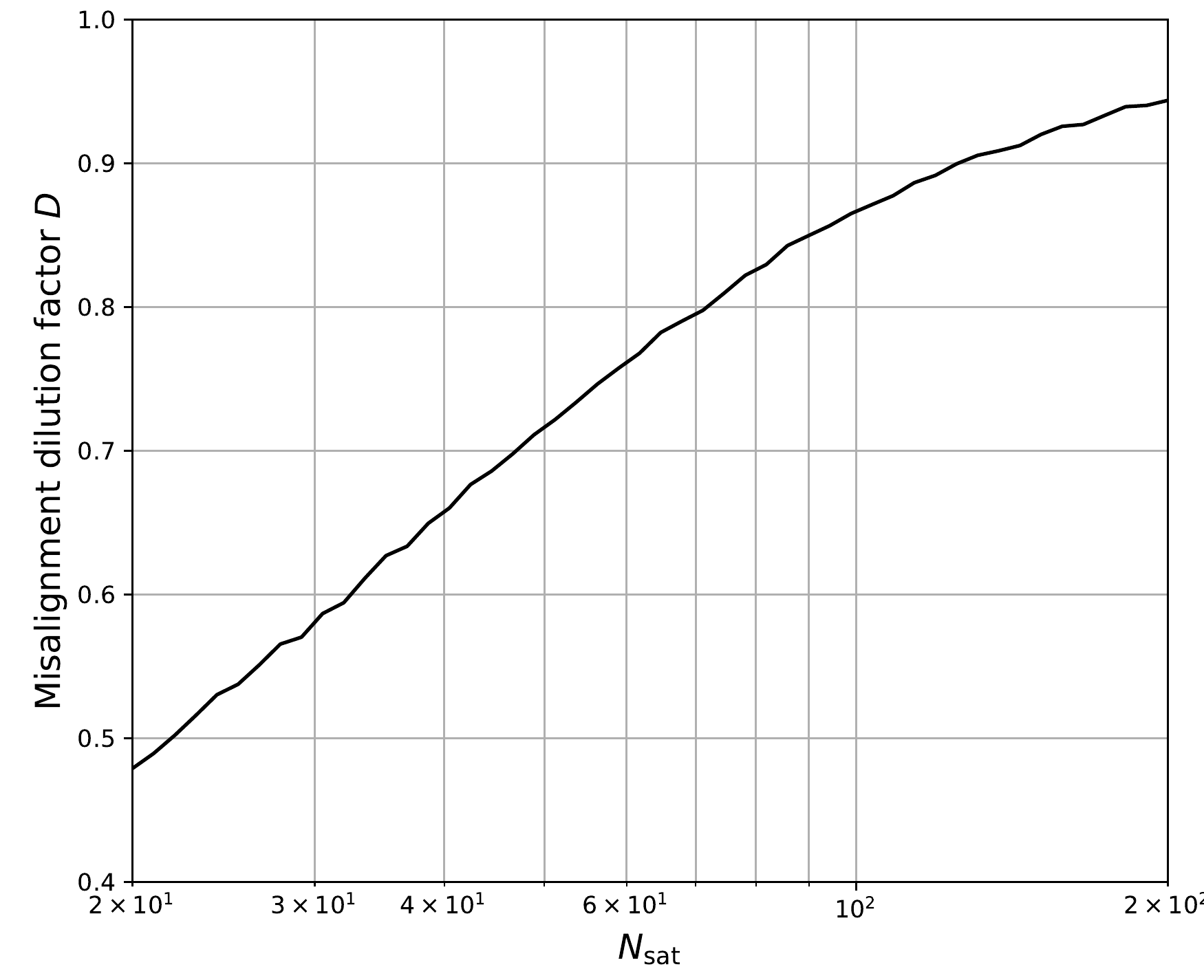}}
\caption{Dilution factor from satellite-halo misalignment as a function of cluster richness, plotted for our best-fit ellipticity 0.271.
It is calculated by applying Eq.~(\ref{eq:dilute}) to a large sample of monte-carlo clusters.
}
\label{fig:dilution}
\end{figure}

\begin{figure*}
\centering
\resizebox{180mm}{!}{\includegraphics{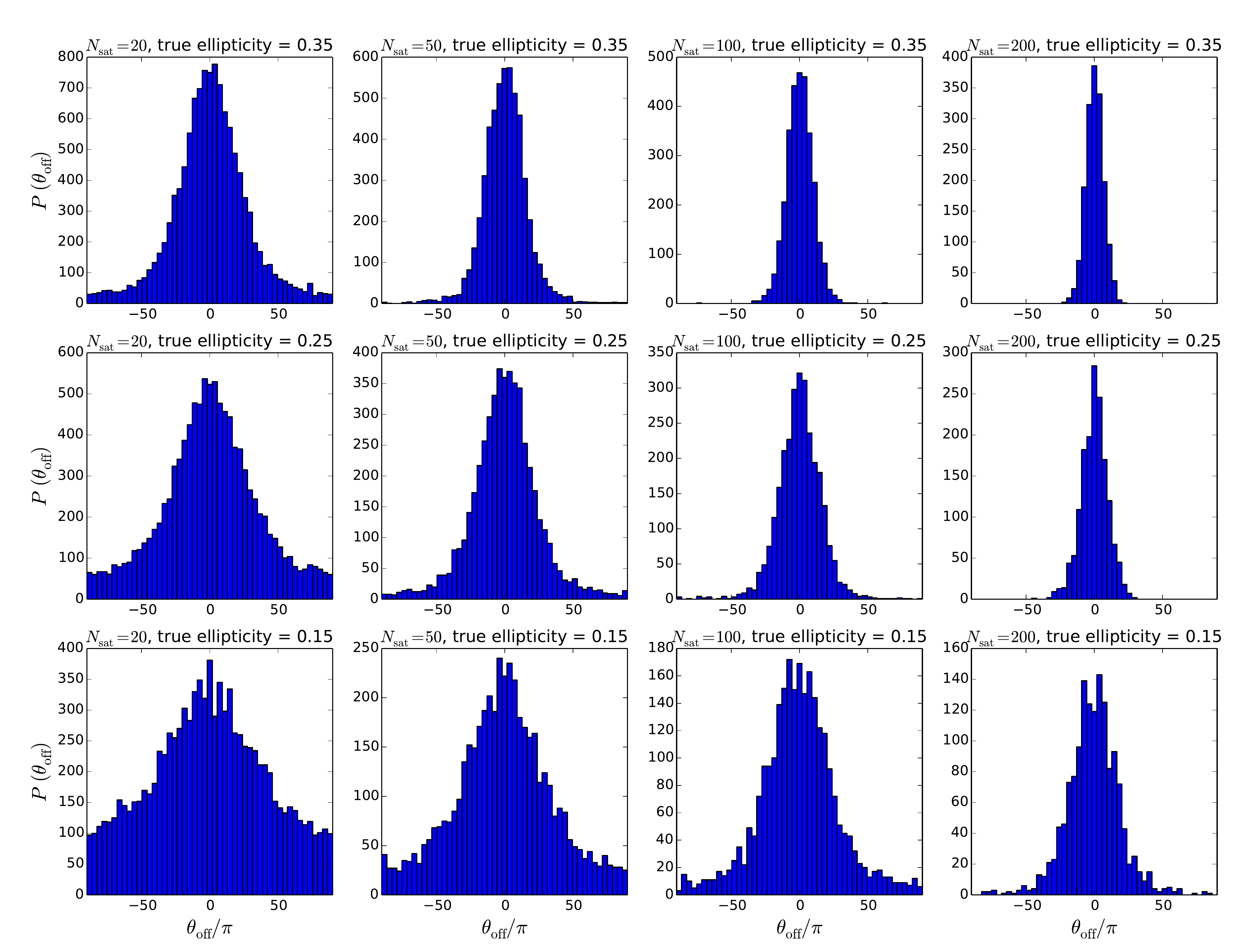}}
\caption{Alignment distributions for various values of $\nsat$ and true ellipticity.
The distribution $P(\theta_{\rm off})$ has the same definition as in \citet{clampitt16}.
}
\label{fig:misalign}
\end{figure*}

The dilution factor described in Sec.~\ref{sec:lensing} is shown in Fig.~\ref{fig:dilution} for ${\rm \epsilon_{true}} = 0.271$, the true mean ellipticity of satellite distribution measured in Sec.~\ref{sec:satellite}.
Since the major axes along which we align the clusters are calculated with satellite distributions, we must derive D assuming the true mean ellipticity to be that of the {\it satellite} distribution. 
Once the value of ellipticity is assumed, we use the same MC simulation as described in Sec.~\ref{sec:satellite} to obtain the dispersion of misalignment angle of major axes of satellite distribution with respect to the true major axes of the halos. From this we calculate the misalignment factor D according to the Eq.~(\ref{eq:dilute}).
For illustration, we also show several alignment distributions for various $N_{\rm sat}$ and $\epsilon$ in Fig.~\ref{fig:misalign}.

%%%%%%%%%%%%%%%
%%%%%%%%%%%%%%%

%%%%%%%%%%%%%%%
%%%%%%%%%%%%%%%

%%%%%%%%%%%%%%%%%%%%%%%%%%%%%%%%%%%%%%%%%%%%%%%%%%

% Don't change these lines
\bsp	% typesetting comment
\label{lastpage}
\end{document}